\documentclass[hidelinks, 12pt]{article}
\usepackage[mathscr]{eucal}
\usepackage{amssymb,latexsym}
\usepackage{verbatim}
\usepackage{graphicx}
\usepackage{color}
\usepackage{amsmath}
\usepackage{amsthm}
\usepackage{enumerate}
\usepackage{framed}
\usepackage{authblk}
\usepackage{setspace}
\usepackage{hyperref}

\definecolor{dgray}{RGB}{90,90,90}
\definecolor{gray}{RGB}{120,120,120}
\definecolor{lgray}{RGB}{150,150,150}

\numberwithin{equation}{section}

\newtheorem{thm}{Theorem}[section]
\newtheorem{lem}[thm]{Lemma}
\newtheorem{Def}[thm]{Definition}
\newtheorem{rem}[thm]{Remark}
\newtheorem{prop}[thm]{Proposition}
\newtheorem{cor}[thm]{Corollary}

\newcommand\bbN{{\mathbb N}}
\newcommand\bbR{{\mathbb R}}

\renewcommand\S{\Sigma}
\newcommand\s{\sigma}
\renewcommand\d{\partial}

\renewcommand\L{\triangle}
\newcommand\D{\nabla}
\newcommand\e{\epsilon}
\renewcommand\b{\beta}

\renewcommand\div{{\rm div}}

\newcommand\ric{{\rm Ric}}

\newcommand\g{\gamma}
\newcommand\8{\infty}
\renewcommand\a{\alpha}

\renewcommand\th{\theta}

\newcommand\tlS{\widetilde S}
\newcommand\tlF{\widetilde F}
\newcommand\tlP{\widetilde P}
\newcommand{\field}[1]{\mathbb{#1}}

\DeclareFontFamily{OT1}{rsfs}{}
\DeclareFontShape{OT1}{rsfs}{m}{n}{ <-7> rsfs5 <7-10> rsfs7 <10->
rsfs10}{} \DeclareMathAlphabet{\mycal}{OT1}{rsfs}{m}{n}
\def\scri{{\mycal I}}%

\newcommand\beq{\begin{equation}}
\newcommand\eeq{\end{equation}}
\newcommand\ben{\begin{enumerate}}
\newcommand\een{\end{enumerate}}
\newcommand\bit{\begin{itemize}}
\newcommand\eit{\end{itemize}}

\newcounter{mnotecount}[section]

\title{
Achronal limits, Lorentzian spheres, and splitting}
\author{Gregory J. Galloway\thanks{This work was partially supported by NSF grants DMS-0708048 and DMS-1313724 (GJG) and by a grant from the Simons Foundation (Grant No.  63943 to G. J. Galloway)} }
\author{Carlos Vega}
\affil{Department of Mathematics
\\ University of Miami }

\begin{document}

\date{}

\maketitle

\begin{abstract} 

In the early 80's S.-T.\! Yau posed  the problem of establishing the rigidity of the Hawking-Penrose singularity theorems.  Approaches to this problem have involved the introduction of Lorentzian Busemann functions and the study of the geometry of their level sets - the horospheres. The regularity theory in the Lorentzian case is considerably more complicated and  less complete than in the Riemannian case.
In this paper we introduce a  broad generalization of the notion of horosphere in Lorentzian geometry and take a completely different (and highly geometric) approach to regularity. These generalized horospheres are defined in terms of {\it achronal limits}, and the improved regularity we obtain is based on regularity properties of achronal boundaries.  
We establish a splitting  result for generalized horospheres, which when specialized to 
{\it Cauchy horospheres} yields new results on the Bartnik splitting conjecture,  a concrete realization of the problem posed by Yau.  Our methods are also applied to spacetimes with positive cosmological constant. We obtain a rigid singularity result for future asymptotically de Sitter spacetimes related to results in \cite{AndGal, zero}.

\end{abstract}

\newpage

\setcounter{tocdepth}{2}

\tableofcontents

\section{Introduction}

In the early 80's S.-T.\! Yau posed the problem of establishing the {\it rigidity} of the Hawking-Penrose singularity theorems.  Approaches to this problem, including the approach advocated by Yau \cite{Yau},  have involved the introduction of Lorentzian Busemann functions and the study of the geometry of their level sets - the horospheres. The regularity theory in the Lorentzian case is considerably more complicated and  less complete than in the Riemannian case.
In this paper we introduce a  broad generalization of the notion of horosphere in Lorentzian geometry and take a completely different (and highly geometric) approach to regularity. These generalized horospheres are defined in terms of {\it achronal limits} (see Section \ref{AL}), and the improved regularity we obtain is based on regularity properties of achronal boundaries as developed by Penrose \cite{Penrose}.
Generalized horospheres are introduced in Section \ref{SHS}, along with the two main classes of examples, {\it Cauchy horospheres} and {\it ray horospheres}.   The latter are closely related to  standard horospheres.   Cauchy horospheres generalize standard horospheres in the context of spacetimes with compact Cauchy surfaces.  Convexity and rigidity properties of generalized horospheres are studied in Section \ref{CR}.   These results are then applied to obtain some splitting results for globally hyperbolic spacetimes.

The present work was motivated in part by the so-called 
{\it Bartnik splitting conjecture} \cite{Bart88} and some recent related developments (see e.g., \cite{Sharif}). The classical Hawking-Penrose singularity theorems (cf., \cite{HE}) establish the existence
of singularities, expressed in terms of incomplete causal geodesics, in large generic classes
of spacetimes.  In \cite{Geroch} Geroch put forth, in rather explicit terms, the conjectural point of view that spatially closed
spacetimes obeying reasonable energy conditions should fail to be singular only under exceptional circumstances.   In the early 80's, Yau formulated this problem in more 
differential geometric terms as noted above.
This  led to the following 
explicit conjecture, stated as Conjecture 2 in Bartnik \cite{Bart88}. 

\bigskip
\noindent
{\bf Conjecture}. 
{\it 
Let $(M,g)$ be a spacetime which contains a
compact Cauchy surface and obeys the strong energy condition,
$Ric(X,X)\ge0$ for all timelike vectors $X$. If $(M,g)$ is timelike geodesically complete,
then $(M,g)$ splits isometrically into the product
$(\bbR \times V,-dt^2\oplus h)$, where $(V,h)$ is a compact Riemannian manifold.}

\bigskip

Thus, according to the conjecture,  such spacetimes $M$ must be singular, except under very special circumstances; see e.g.\  \cite[Chapter 14]{BEE} for further background.  The conjecture has been proven subject to the addition of   a `no observer horizon' type condition, which has taken various forms (see  e.g., \cite{Gal84, Bart88, EschGal}).  In Section~\ref{CR} we prove a splitting result for generalized horospheres, see Theorem \ref{horosplit}.  When specialized to Cauchy horospheres, this leads to a proof of the Bartnik conjecture provided that, in addition, a certain  `max-min' condition associated to a given Cauchy surface $S$ holds, see Definition \ref{MMdef} in Section \ref{horoapps}. The splitting is shown to occur along the Cauchy horosphere 
$S^-_{\infty}(S)$.   Within the class of timelike geodesically complete spacetimes with compact Cauchy surfaces, we show that this max-min condition is implied by the so-called $S$-ray condition introduced in \cite{EschGal}.   In fact the $S$-ray condition is  a strictly stronger condition within this class, as can be  seen in de Sitter space, where, due to the presence of observer horizons, the $S$-ray condition fails, but where the max-min condition holds.  
 An alternative approach to Theorem \ref{horosplit}, in the case of a Cauchy horosphere, is considered in 
Section~\ref{lines}, by way of the Lorentzian splitting theorem.   In Section 6 we obtain a 
rigid singularity theorem for spacetimes with positive cosmological constant, which is related to some results in \cite{AndGal} (see also \cite{zero}), but which does not assume the existence of a future conformal completion.  The proof of this theorem involves the introduction of the notion of `limit mean curvature'.

\section{Achronal Limits}\label{AL} 

With regard to causal theoretic notions  discussed here and elsewhere, we refer the reader to the standard references \cite{BEE, HE, Penrose, ON}.
Let $(M^{n+1},g)$ be a spacetime, i.e., a connected, time-oriented Lorentzian manifold, with 
$n \ge 1$.  We recall here some basic facts about {\it achronal boundaries}. 

A set $F \subset M$ is called a {\it future set} if $F = I^+(S)$ for some set $S \subset M$.  It follows  that  $F$ is a future set if and only if  $I^+(F) = F$. A {\it past set} is defined time-dually. Note that future and past sets are necessarily open. An achronal boundary is the boundary (assumed to be non-empty)  of a future or past set, i.e., a non-empty set of the form $A = \partial I^\pm(S)$, for some subset $S \subset M$. Achronal boundaries have many nice structural properties.  In particular, an achronal boundary $A \subset M$  is in general an  edgeless achronal $C^0$ hypersurface of $M$ (cf. \cite[Lemma 3.17, Corollary 5.9]{Penrose}).

We will rely in an essential way on the following facts about achronal boundaries, cf., \cite[Proposition 3.15]{Penrose}.

\begin{prop} [Achronal Decomposition]\label{decomp}
Let $A$ be a (nonempty) achronal boun-dary.   Then we have the following. 
\ben
\item[(1)] There exists a unique future set $F$ such that $\d F = A$ and  a unique past set $P$ such that $\d P = A$.  Then also,  $I^+(A) \subset F$ and $I^-(A) \subset P$.
\item[(2)] The sets  $P$, $F$ and $A$ are mutually disjoint and,
\beq\label{eqdecomp}
M = P \cup A \cup F\,.
\eeq
Further, any  curve from $P$ to $F$ must meet $A$ (at a unique point if it is timelike).
\een
\end{prop}

\smallskip
\proof[Remark on the proof.]  The statement above is a slight refinement  of
the statement  of Proposition 3.15 in \cite{Penrose}.  Suppose, for example, $A = \d F$, where $F$ is a future set.  Then, as in \cite{Penrose}, $P = I^-(M\setminus F)$ is a past set with $A = \d P$, and the decomposition \eqref{eqdecomp} holds.  If $A = \d F'$, for some other future set $F'$, then in a similar fashion we are led to the decomposition $M = P' \cup A \cup F'$.  But Proposition 3.15 in \cite{Penrose} then implies that $F' = F$. \qed

\medskip
Simple examples show that the inclusions  $I^-(A) \subset P$ and $I^+(A) \subset F$ may be strict.
We say that an achronal boundary is {\it past proper} provided $I^-(A) = P$. It follows from the uniqueness of $P$ that $A$ is past proper if and only if $\d I^-(A) = A$.  Time-dually, an achronal boundary $A$ is {\it future proper} provided $I^+(A) = F$ (or, equivalently, provided $\d I^+(A) = A$).   We have the following basic lemma whose proof is left to the reader.

\begin{lem}\label{achronalmono} 
Let $A$ and $B$ be achronal boundaries with associated unique past and future sets $\{P_A, F_A\}$ and  $\{P_B, F_B\}$, respectively.  Then the following hold.
\ben
\item [(1)] $P_A \subset P_B$ if and only if $F_B \subset F_A$.
\item [(2)] If $A$ and $B$ are past proper, then $P_A \subset P_B$ iff $J^-(A) \subset J^-(B)$ iff $A \subset J^-(B)$.
\een
\end{lem}

\smallskip
We will say that a sequence of achronal boundaries $\{A_k\}$ is \emph{monotonic} if either $\{P_k\}$ is increasing (i.e., $P_k \subset P_{k+1}$  for all $k$) or 
$\{F_k\}$ is increasing.

\begin{Def}[Achronal Limits]\label{alimits}  
Let $\{A_k\}$ be a sequence of achronal boundaries, and, for each $k$, let $P_k$ and $F_k$ be the unique past and future sets such that  $A_k = \d P_k = \d F_k$.
\ben
\item[(1)] If the sequence $\{P_k\}$ is increasing then the \emph{future achronal limit} $A_\infty$ of  
$\{A_k\}$ is defined as,
\beq
A_\infty = \partial  \bigg( \bigcup_{k} P_k\bigg)   \,.
\eeq
\item[(2)] If the sequence $\{F_k\}$ is increasing then the \emph{past achronal limit} $A_\infty$ of  
$\{A_k\}$ is defined as,
\beq
A_\infty = \partial  \bigg( \bigcup_{k} F_k\bigg)   \,.
\eeq
\een
\end{Def}

\smallskip
Since an arbitrary union of past (resp. future) sets is a past (resp. future) set,  an achronal limit (if non-empty) is an achronal boundary,  and hence is itself an edgeless achronal $C^0$ hypersurface in $M$.   

\begin{rem}\label{pastproper}
{\rm  Suppose, as will be the case in examples later, that $\{A_k\}$ is a sequence of past proper achronal boundaries, i.e.,  $I^-(A_k) = P_k$ for all $k$.  Then, by Lemma~\ref{achronalmono}, $\{P_k\}$ is increasing  if and only if   $\{J^-(A_k)\}$ is increasing if and only if $A_k \subset J^-(A_{k+1})$ for all $k$. Similarly, $\{F_k\}$ is increasing, or equivalently 
$\{P_k\}$ is decreasing,  if and only if $\{J^-(A_{k})\}$ is decreasing if and only if $A_{k+1} \subset J^-(A_{k}$) for all $k$.  
}
\end{rem}

We now show that an achronal limit $A_{\infty}$ is, in a sense made precise below, the sequential limit of the $A_k$'s.
A sequence of points $\{x_k\}$ is {\it future increasing} if $x_k \le x_{k+1}$ for all $k$. If such a sequence converges to $x \in M$, we call $x$ the {\it (future) causal limit} of $\{x_k\}$.  {\it Past increasing} sequences of points, and their limits, are understood time-dually. 

\smallskip
\begin{prop} [Sequential Characterization of Achronal Limits] \label{sequential}
Let $A_\infty$ be the future (resp. past) achronal limit of a sequence of achronal boundaries $\{A_k\}$. Then any limit point of a sequence $a_k \in A_k$ is contained in $A_{\infty}$.  Moreover, fixing any point $a \in A_\infty$ and any timelike curve $\a$ through $a$, $a$ is the future (resp. past) causal limit of the sequence $a_k = \a \cap A_k$, (for $k$ sufficiently large).
\end{prop}

\proof
For definiteness, assume $A_\infty$ is a future limit. By Definition \ref{alimits}, $\{P_k\}$ is increasing and $A_{\infty} = \d P_{\infty}$, $P_{\infty} = \cup_k P_k$.

Suppose $a \in M$ is a limit point of a sequence $a_k \in A_k$, with $a_{k_j} \to a$. Let $U$ be any neighborhood of $a$. For large $j$, we have $a_{k_j} \in U$. Then for large $j$, since $U$ meets $A_{k_j} = \d P_{k_j}$, it meets $P_{k_j}$ and hence also $\cup_k P_k = P_{\infty}$. Also, $U$ intersects $I^+(a)$ at some point $y$, and since $I^-(y)$ is open and contains $a$, it contains $a_{k_j}$ for all sufficiently large $j$.  Thus, $y \in I^+(A_{k_j}) \subset F_{k_j}$ for all large $j$, and consequently, $y \notin \cup_k P_k = P_{\infty}$. It follows that $a \in \d P_{\infty} = A_{\infty}$. Hence, $A_\infty$ contains all limit points of sequences $a_k \in A_k$.

Now let $a \in A_\infty$ and let $\alpha : I \to M$ be any future pointing timelike curve with $0 \in I$ and $\alpha(0) = a$. Fix $T > 0$ with $-T \in I$. We have $\alpha|_{[-T, 0)} \subset I^-(A_\infty) \subset P_{\infty}  = \cup_k P_k$ and $a  \not \in P_k$. Thus, for all sufficiently large $k$, $\alpha$ is a timelike curve from $\alpha(-T) \in P_k$ to $\alpha(0) = a \in A_k \cup F_k$. It follows then from Proposition~\ref{decomp} that for each sufficiently large $k$, there is a unique $t_k \in (-T, 0]$ such that $a_k := \alpha(t_k) \in A_k$. The fact that $\{P_k\}$ is  increasing implies that $\{t_k\}$ must be (weakly) increasing in $k$, and hence that $\{a_k\}$ is future increasing. Suppose that $t_k \not \to 0$, i.e., that $t_k \le 2\delta < 0$. Then $\alpha(\delta) \in I^-(A_\infty) \subset P_{\infty}$,  and hence $\alpha(\delta) \in P_k$ for large $k$.  On the other hand,  $\alpha(\delta) \in I^+(A_k) \subset F_k$ for large $k$, which is not possible since $P_k \cap F_k = \emptyset$.
So we have $t_k \to 0$, and thus $a_k = \alpha(t_k) \to \alpha(0) = a$.
\qed

\smallskip
\begin{rem}
{\rm Proposition \ref{sequential} is closely related to the notion of {\it Hausdorff closed limits}; see, for example, \cite[Section 4.3]{Papadopoulos}.  In fact, it follows from Proposition \ref{sequential} that the Hausdorff closed limit of a monotonic sequence of achronal boundaries $\{A_k\}$ exists and coincides with its achronal limit, and hence we may write
$A_\infty = \lim  A_k $.}
\end{rem}

\section{Spheres and horospheres}
\label{SHS}

Throughout this section we shall assume spacetime $M$ is globally hyperbolic.
Recall, this means that 
$M$ is strongly causal and that all sets of the form $J^+(p) \cap J^-(q)$ are compact.  By a Cauchy surface for $M$ we shall mean an achronal set $S \subset M$ that is met by every inextendible causal curve in 
$M$.   A Cauchy surface for $M$ is,  in particular, a (past and future proper) achronal boundary, and hence is an edgeless achronal $C^0$ hypersurface in~$M$.  The following facts are well-known, cf., \cite{HE, Penrose, ON}:  
\ben
\item[(1)] $S$ is a Cauchy surface for $M$ if and only if $S$ is achronal and $D(S) = M$, or equivalently, $H(S) = \emptyset$, where $D(S) = D^+(S) \cup  D^-(S)$ is the domain of dependence of $S$ and $H(S) = H^+(S) \cup  H^-(S)$ is the Cauchy horizon of $S$.
\item[(2)] $M$ is globally hyperbolic if and only if it admits a Cauchy surface $S$.
\een

\subsection{Lorentzian length and  distance} 

We denote the Lorentzian distance function by $d$; letting $\Omega^c_{p,q}$ be the set of future directed causal curves from $p$ to $q$, recall that,

\beq
 d(p,q) =
 \begin{cases}
\, \sup \{L(\gamma) : \gamma \in \Omega^c_{p,q}\},   & \quad q \in J^+(p)\\ 
\, 0, & \quad q \notin J^+(p)  \,.
\end{cases}
\eeq

\smallskip
A future causal curve segment $\alpha$ from $p$ to $q \in J^+(p)$ is {\it maximal} if $d(p,q) = L(\alpha)$, i.e., there is no longer causal curve from $p$ to $q$. Note that this implies $\alpha$ realizes the distance between any two of its points. A future {\it ray} is a future-inextendible curve $\gamma : [a, b) \to M$, where $b \in (a,\infty]$, each segment of which is maximal.

As $M$ is taken to be globally hyperbolic, we have the standard fact that $d$ is finite and continuous on $M \times M$. Furthermore, any two causally related points are joined by a maximal causal geodesic segment.

For any subset $S \subset M$, we define the distance from $S$ to $q \in M$ by
\beq
 d(S,q) =
 \begin{cases}
\, \sup \{ d(x,q) : x \in S\} & \quad q \in J^+(S)\\
\, 0, & \quad q \notin J^+(S)
\end{cases}
\eeq

A future causal curve segment $\alpha$ from $S$ to $q \in J^+(S)$ is called a future {\it maximal $S$-segment}  if $d(S, q) = L(\alpha)$. Again, this implies that $\alpha$ realizes the distance from $S$ to any of its points. A future {\it $S$-ray} is a future-inextendible curve $\gamma : [a,b) \to M$ from $\gamma(a) \in S$, such that each initial segment 
$\gamma |_{[a,c]}$, $c \in (a,b)$, is a maximal $S$-segment.  The distance to $S$ from a point $q$, 
$d(q,S)$, is defined in a similar fashion, and analogous terminology is used.  

\subsubsection{Causal completeness and boundedness}
We shall make use of the notion of causal completeness introduced in \cite{Gcambphil}; see also~\cite{Grant, Treude}.

\begin{Def} {\rm A subset $S \subset M$ is said to be \emph{future causally complete} if for all $p \in J^+(S)$, the closure in $S$ of $J^-(p) \cap S$ is compact. \emph{Past causal completeness} is defined time-dually.}
\end{Def}
We note that compact sets and Cauchy surfaces are  past and future causally complete. 
In general, if $S$ is either past or future causally complete, then $S$ is necessarily closed:

\begin{lem} \label{CCisclosed} If $S$ is future causally complete, then $S$ is closed.
\end{lem}
\proof Consider a sequence $\{s_k\} \subset S$, with $s_k \to x$. Let $y \in I^+(x)$. Then for all large $k$, $s_k \in J^-(y) \cap S$, and, in particular, $y \in J^+(S)$. Since the closure in $S$ of $J^-(y) \cap S$ is compact, the sequence $\{s_k\}$ has a convergent subsequence in $S$, which must in fact converge to 
$x$. Hence, $x \in S$. \qed
 
\medskip
We summarize several other basic properties of causal completeness, cf., \cite{Treude, Vega}.

\medskip
\begin{lem} \label{GHCC} Let $S$ be a subset of a  globally hyperbolic  spacetime $M$.
\ben

\vspace{-.5pc}
\item [(1)] If $S$ is future causally complete then $J^+(S)$ is closed.

\vspace{-.3pc}
\item [(2)] The following are equivalent for $S$ closed.

\vspace{-.5pc}
\ben
\item $S$ is future causally complete.
\item $J^-(p) \cap S$ is compact for all $p \in J^+(S)$.
\item $J^-(p) \cap J^+(S)$ is compact for all $p \in J^+(S)$.
\een

\vspace{-.8pc}
\item [(3)] If $C$ is future causally complete and $S \subset J^+(C)$ is closed, then $S$ is future causally complete.
\een
\end{lem}
\proof[Remarks on the proof] (1) is proved in a manner similar to Lemma \ref{CCisclosed}.  For (2), the equivalence $(a)\iff(b)$ follows readily from the fact that $S$ is closed. The equivalence $(b)\iff(c)$ follows from the set relations, $J^-(p) \cap S \subset J^-(p) \cap J^+(S) = J^-(p) \cap J^+(J^-(p) \cap S)$. For (3), note that if $x \in J^+(S)$, then $J^-(x) \cap S$ is a closed subset of the compact set $J^-(x) \cap J^+(C)$. \qed

\medskip
The following is fundamental to our treatment of generalized spheres. 

\begin{lem}\label{CCdist} Let $M$ be globally hyperbolic. If $S$ is future causally complete, then $x \to d(S,x)$ is finite-valued and continuous on $M$, and given any $q \in J^+(S)$, there is a maximal future $S$-segment $\a$ from $S$ to $q$, i.e., $L(\a) = d(S,q)$.
\end{lem} 
 
\proof  Let $q \in J^+(S)$, and note that $J^-(q) \cap S$ is compact. Let $x_k \in J^-(q) \cap S$ such that $d(x_k, q) \to d(S,q)$. Then $\{x_k\}$ has a limit point $p_0 \in S$, and by continuity of $d$ on $M \times M$, we have $d(p_0, q) = \lim_{j \to \infty} d(x_{k_j}, q) = d(S,q)$. In particular, $d(S,q) < \infty$. Furthermore, by global hyperbolicity,  $q \in J^+(p_0)$, hence, $p_0$ is joined to $q$ by a maximal causal geodesic segment $\a$, which must also be maximal as an $S$-segment, $L(\a) = d(p_0,q) = d(S,q)$. For continuity, note that, since $J^+(S)$ is closed, $d(S, \cdot)$ is continuous on the open set $M \setminus J^+(S)$, where it vanishes identically. Hence it remains to show continuity at $q \in J^+(S)$. Note that for this, it suffices to show that for any sequence $q_k \to q$, we have $\lim_{j \to \infty} d(S, q_{k_j}) =d(S,q)$, for some subsequence $\{q_{k_j}\}$, (since this would apply to a supremum-realizing sequence as well as an infimum-realizing sequence). Fix $q_+ \in I^+(q) \subset J^+(S)$. Then $J^-(q_+) \cap S$ is compact. For all large $k$, we have $q_k \in J^-(q_+)$ and hence $J^-(q_k) \cap S \subset J^-(q_+) \cap S$. Let $p_k \in J^-(q_+) \cap S$ with $d(p_k, q_k) = d(S, q_k)$, where $p_k$ is chosen arbitrarily for any $q_k \not \in J^+(S)$. By compactness of $J^-(q_+) \cap S$, $\{p_k\}$ has a subsequence $\{p_{k_j}\}$ converging to some $p_\infty \in J^-(q_+) \cap S$. By continuity on $M \times M$, we have $\lim_{j \to \infty} d(S, q_{k_j}) = \lim_{j \to \infty} d(p_{k_j}, q_{k_j}) = d(p_\infty, q) \le d(S, q) = d(p_0, q)$. On the other hand, we have $d(p_0, q_{k_j}) \le d(S, q_{k_j})$, the limit of which gives $d(p_0, q) \le \lim_{j \to \8} d(S, q_{k_j})$. Hence $\lim_{j \to \infty} d(S, q_{k_j}) = d(p_0, q) = d(S, q)$. \qed

 \medskip 
In view of Lemma \ref{CCdist}, causal completeness is what we will typically demand when considering distance from a set.  Causal completeness has been used in \cite{Grant, Treude} for  similar purposes.

\smallskip

A past or future causally complete set $S$,  which is also achronal and edgeless, enjoys some additional properties.

\begin{lem} \label{PPifPCC} Let $M$ be globally hyperbolic and suppose $\emptyset \ne S \subset M$ is achronal and edgeless. If $S$ is past causally complete, then $S = \d I^-(S)$, i.e., $S$ is a past proper achronal boundary.
\end{lem}

\proof By achronality, $S \cap I^-(S) = \emptyset$. It follows that $S \subset \d I^-(S)$. We show that 
$\d I^-(S) \subset S$. Suppose otherwise that $x \in \d I^-(S) \setminus S$.  By Theorem 3.20 in \cite{Penrose}, 
$x$ is the past endpoint of a null geodesic $\eta \subset \d I^-(S)$ which is either future inextendible in $M$ or else has a future  endpoint on $S$.  Since $J^-(S)$ is closed, we have 
$\eta \subset J^+(x) \cap \d I^-(S) \subset J^+(x) \cap J^-(S)$. If $\eta$ were future inextendible, being imprisoned in the compact set $J^+(x) \cap J^-(S)$ would imply a strong causality violation. Hence, 
$\eta$ must have a future endpoint $y \in S$; we may assume $\eta \cap S = \{y\}$.  Then since $S$ is an achronal  $C^0$ hypersurface and $y$ is the only point of $\eta$ to meet $S$, there will be a point on $\eta$ near $y$ in the timelike past of $S$.  This implies that $x \in I^-(S)$, which is a contradiction.\qed  

\medskip
In a somewhat similar fashion we obtain the following.

\begin{lem} \label{PCCiffPCS} Let $M$ be globally hyperbolic and suppose $\emptyset \ne S \subset M$ is achronal and edgeless. Then $S$ is past causally complete if and only if $S$ is a past Cauchy surface,  $H^-(S) = \emptyset$. 
\end{lem}

\proof Suppose $H^-(S) = \emptyset$. Then $D^-(S) = J^-(S)$, from which it follows that $J^+(p) \cap S$ is compact for all $p$ in $J^-(S)$.  Hence, $S$ is past causally complete.  Now suppose that $S$ is past causally complete, and that there exists a point $p \in H^-(S)$. By Theorem 5.12 in \cite{Penrose}, $p$ is the past endpoint of a null geodesic $\eta \subset H^-(S)$ which is future-inextendible in $M$. But this 
again leads to a strong causality violation,
since $\eta \subset J^+(p) \cap H^-(S) \subset J^+(p) \cap J^-(S)$. Hence, $H^-(S) = \emptyset$. \qed

\medskip  In globally hyperbolic spacetimes, causal completeness is closely related to the notion of `causal boundedness', which will be used later to ``causally control" subsets of $M$.

\begin{Def} {\rm We say a subset $A \subset M$ is \emph{future bounded} if there is a Cauchy surface $S$ in $M$ such that $A \subset J^-(S)$. Past boundedness is defined time-dually.}
\end{Def}

\begin{lem}\label{PCCiffFB} Let $M$ be globally hyperbolic and $\emptyset \ne S \subset M$ closed. Then $S$ is past causally complete  if and only if $S$ is  future bounded.
\end{lem} 

\proof First suppose $S$ is future bounded by a Cauchy surface $\S$, i.e., $S \subset J^-(\S)$. Since $\S$ is past causally complete, then $S$ is past causally complete, by Lemma \ref{GHCC}.

Now suppose $S$ is past causally complete. Let $A = \d I^-(S)$. We first note that $A$ is nonempty.  Otherwise, $M = I^-(S) = J^-(S)$. But then the causal future of any point would be compact, by Lemma \ref{GHCC}, which would imply a strong causality violation. Hence, $A$ is a nonempty edgeless achronal  set.   Using \cite[Theorem 3.20]{Penrose} as in the proof of Lemma \ref{PPifPCC}, we have, $A \subset J^-(S)$.   Hence, by Lemma~\ref{GHCC}, $A$ is also past causally complete.   Then by Lemma \ref{PCCiffPCS}, $A$ is a past Cauchy surface, i.e., $H^-(A) = \emptyset$ and $D^-(A) = J^-(A)$. 

Let  $\widetilde{M} = M - J^-(A)$.  Since, in particular $J^-(A)$ is closed, one readily verifies that $\widetilde M$ is a  (connected) globally hyperbolic spacetime in its own right. As such it admits a Cauchy surface $\S$.  We claim that $\S$ is a Cauchy surface for $M$.  Let $\b : (-\8,\8) \to M$ be a future directed, inextendible causal curve in $M$. If $\b \subset \widetilde{M}$, then $\b$ must meet $\S$. On the other hand if $\b$ meets $J^-(A) = D^-(A)$ at some point $\b(t)$, say, then $\b|_{[t,\8)}$  must intersect  $A$ and enter into  $\widetilde{M}$. But the portion of $\b$ in $\widetilde{M}$ must then meet $\S$.  Hence, $\S$ is a Cauchy surface for $M$.   

Finally, fix any $x \in S$. Let $\a$ be any future-inextendible causal curve from $x \in S$. Since $J^+(x) \cap J^-(S)$ is compact, $\a$ must eventually leave $J^-(A) \subset J^-(S)$, and enter $\widetilde{M}$.   Hence, the future end of $\a$ meets $\S$, which means $x \in J^-(\S)$. Since $x \in S$ was arbitrary, we have $S \subset J^-(\S)$. \qed

\subsubsection{Further facts}
We will need the following fact later.

\begin{lem}\label{nullray} 

Let $S$ be an achronal boundary and $\eta : [0, b] \to M$, $\eta(0) \in S$, a future directed  null geodesic.   If $\eta$ is $S$-maximal then $\eta \subset S$.  Hence, a null $S$-ray from an achronal boundary $S$ is necessarily contained in $S$.
\end{lem}

\proof Letting $P$ and $F$ be the unique past and future sets, respectively, as in Proposition~\ref{decomp}, we have $\eta \subset J^+(S) \subset \overline{I^+(S)} \subset S \cup F$. Suppose that $\eta(c) \in F$ for some $c \in (0,b]$. Since $\eta(0) \in S$, we have $I^-(\eta(0)) \subset P$. Consequently, there is a timelike curve from $P$ to $\eta(c) \in F$, and by the separating property of achronal boundaries, this curve must meet $S$. But this means $\eta(c) \in I^+(S)$, which implies $L(\eta|_{[0,c]}) = 0 < d(S,\eta(c))$, contradicting the maximality of $\eta$ as an $S$-segment.\qed

\medskip
The limit curve lemma is a basic tool in Lorentzian geometry.  We will make use of it in the following form, cf. \cite{BEE, Gcambphil}.

\begin{lem} 
 Fix a complete Riemannian metric $h$ on $M$. Let $\alpha_k : [0, \infty) \to M$ be a sequence of future-inextendible causal curves, parameterized with respect to $h$-arc length. If $\{\alpha_k(0)\}$ has a limit point $p$, there is a future-inextendible (continuous) causal curve $\alpha : [0, \infty) \to M$ with 
 $\alpha(0) = p$, and a subsequence $\{\alpha_{k_j}\}$ converging uniformly to $\alpha$, with respect to $h$, on compact parameter intervals.
\end{lem}

The limit curve lemma will be used in conjunction with the following standard fact 
(see, e.g. \cite[Theorem 7.5]{Penrose}.

\begin{prop}The Lorentzian arc length functional is upper semicontinuous with respect to the topology of uniform convergence on compact subsets, i.e., if a sequence of causal curves $\alpha_k : [a, b] \to M$ converges uniformly to the causal curve $\alpha : [a, b] \to M$, then
\beq
\limsup_{k \to \infty} L(\alpha_k) \le L(\alpha)
\eeq 
\end{prop}

\medskip
\subsection{Generalized spheres}

Given $r > 0$ and a non-empty past causally complete set $C$, the (generalized) {\it past sphere
of  radius $r$ and  center} $C$ is the set,
\beq
 S^-_r(C) := \{x \; : \; d(x, C) = r\}  \,.
\eeq 

\begin{lem}\label{spheres}
If non-empty, $S^-_r(C)$ is an acausal past proper achronal boundary.
\end{lem}
\proof  Let $S : = S^-_r(C) \ne \emptyset$. We first observe that $S$ is acausal.  
Let $y \in S$ and, by the past causal completeness of $C$, let $\beta$ be a maximal timelike segment from $y$ to $C$ of length $r$. If $z \in J^-(y) - \{y\}$, then by ``cutting the corner" in a neighborhood of $y$ if necessary, one can produce a causal curve from $z$ to $C$ of length strictly greater than $r$. Hence $z \not \in S^-_r(C) = S$. This shows that $S$ is acausal.
To show that $S = \partial I^-(S)$, let $x \in \partial I^-(S)$ and let $\alpha : [-T,0] \to M$ be a future timelike curve segment ending at $\alpha(0) = x$. Then $\alpha |_{[-T,0)} \subset I^-(S)$. Consequently, we have $d(\alpha(-t), C) > r$ for all $t > 0$ and by continuity, $d(x,C) \ge r$. If $d(x, C) > r$, then use a past maximal $C$-segment from $x$ to $C$ to see that $x \in I^-(S)$, a contradiction. Thus, $d(x,C) = r$, i.e., $x \in S$. 
This shows $\d I^-(S) \subset S$. The inclusion $S \subset \partial I^-(S)$ follows from the achronality of $S$.
\qed 

\smallskip
\begin{lem}\label{maxseg}
Let $S$ be a past sphere of radius $r$. Then $S$ admits a maximal future `{\it  radial}' $S$-segment of length $r$ from each point.
\end{lem}
\proof If $S = S^-_r(C)$ for some past causally complete $C \subset M$, then  by the time-dual of Lemma \ref{CCdist}  each $x \in S$ is joined to $C$ by a maximal $C$-segment of length $r$. By definition of $S^-_r(C)$, this segment is necessarily maximal as an $S$-segment.  \qed

\smallskip
\begin{lem}\label{equidist}
Let $C \subset M$ be past causally complete. Then the radius is additive in the  sense that,
$$
S^-_a(S^-_r(C)) = S^-_{r+a}(C)  \,.
$$

\smallskip
\noindent
Consequently,  for $s > r > 0$, we have, $d(S^-_s(C), S^-_r(C)) = s-r$ (provided $S^-_s(C) \ne \emptyset$).
\end{lem}

\proof By Lemma \ref{GHCC}, each sphere $S^-_r(C)$ is past causally complete. Hence, $S^-_a(S^-_r(C))$ is well-defined. To show that the radius is additive, first let $x \in S^-_a(S^-_r(C))$. By Lemma \ref{CCdist}, $x$ is joined to some $y \in S^-_r(C)$ by a timelike segment of length $a$. As $y$ is similarly joined to $C$ by a segment of length $r$, we have $d(x, C) \ge r +a$. Then, letting $\a$ be a maximal future $C$-segment from $x$ to $C$, $\a$ must pass through $S^-_r(C)$. The portion of $\a$ before $S^-_r(C)$ is bounded in length by $a$ and the portion after by $r$, thus $d(x, C) \le r +a$, so $d(x, C) = r +a$, i.e., $x \in S^-_{r+a}(C)$. Now let $x \in S^-_{r+a}(C)$. Then there is a maximal $C$-segment $\a$ from $x$ to $C$ of length $r+a$. As any portion of $\a$  ending at $C$ must also be $C$-maximal, the point $x^\prime \in \a$ from which the remaining portion of $\a$ has length $r$ is a maximal $C$-segment of length $r$, and hence $x^\prime \in S^-_r(C)$, so $d(x, S^-_r(C)) \ge d(x, x^\prime) = a$. But, of course, $d(x, S^-_r(C)) \le a$, since otherwise, one could produce a curve from $x$ to $C$ of length greater than $r+a$. Thus, $d(x, S^-_r(C)) = a$, i.e., $x \in S^-_a(S^-_r(C))$. Hence, $S^-_a(S^-_r(C)) = S^-_{r+a}(C)$. From this we see that for $s > r$, $d(S^-_s(C), S^-_r(C)) = d(S^-_{s-r}(S^-_r(C)), S^-_r(C)) = s-r$.
\qed

\medskip
Future spheres $S^+_r(C)$,  where $C$ is future causally complete, are defined in a similar fashion, namely, $ S^+_r(C) := \{x \; : \; d(C, x) = r\}$.  If $C$ consists of a single point or is a Cauchy surface, then, as noted earlier, $C$ is both future and past causally complete and we call $S^\pm_r(C)$ a {\it point sphere} or {\it Cauchy sphere}, respectively.

\smallskip
The following fact is basic to our development (see also \cite{GalBanach}).

\begin{lem}\label{Cauchyspheres}
Suppose $M$ is future (resp.\ past) timelike geodesically complete. Then future (resp.\ past) Cauchy spheres from a compact Cauchy surface are (compact) Cauchy surfaces.
\end{lem}
\proof
As the proof uses standard arguments, we shall be brief.  Let $C$ be a compact Cauchy surface, and assume $M$ is future timelike geodesically complete.
We show that  $S^+_r(C)$ is compact.  By continuity of the distance function, it is a closed set.
If it is not compact, then, fixing any complete Riemannian metric $h$ on $M$,  there is a sequence of points $x_k \in  S^+_r(C)$ such that the $h$-distance of $x_k$ to $C$ tends  to infinity.
For each $k$, there is a maximal $C$-segment $\a_k:[0,t_k] \to M$ (parameterized with respect to $h$-arc length), from $a_k \in C$ to $x_k$, with $t_k \to \infty$. By compactness of $C$, 
$a_k$ has a limit point $a \in C$. Passing to a subsequence if necessary, the $\a_k$'s converge   to a  future-inextendible causal limit curve $\a : [0,\infty) \to M$.  As the limit of $C$-maximal segments, $\a$ is a $C$-ray.  
Since $\a$ must enter the timelike future of $C$, it is timelike.
As it is a geodesic, the completeness assumption ensures that $\a$ meets $S^+_r(C)$ at some point $\alpha(t_0)$.  By construction of $\a$, $\a_k(t_0+1)$ (defined for sufficiently large $k$) converges
to $\a(t_0+1) \in I^+(S^+_r(C))$.  Hence $\a_k(t_0+1) \in I^+(S^+_r(C))$ for large $k$, but, since 
$\a_k$ ends on $S^+_r(C)$, this  contradicts the achronality of $S^+_r(C)$.

Thus, $S^+_r(C)$ must in fact be compact.  But a compact edgeless achronal set in a globally hyperbolic spacetime is easily seen to be a Cauchy surface (for example by showing that $H(S) = \emptyset$).\qed

\subsection{Generalized horospheres}\label{genhoro}  

Let $\{S^-_k = S^-_{r_k}(C_k)\}$ be a sequence of (nonempty) past spheres.  By Lemma \ref{spheres},  each $S^-_k$ is a past proper achronal boundary, and hence has unique associated past and future sets, $P^-_k$ and $F^-_k$, as in Proposition \ref{decomp}, with $P^-_k = I^-(S_k^-)$. Recall that
we say $\{S^-_k\}$ is monotonic if either $\{P^-_k\}$ is increasing or $\{F^-_k\}$ is increasing.

\smallskip
\begin{Def} {\rm  Let $\{S_k^- = S_{r_k}^-(C_k)\}$ be a monotonic sequence of past spheres with radii $r_k \to \infty$.
\ben 
\item[(1)]  If $\{P_k^-\}$ is an increasing sequence,  we obtain the future achronal limit,
\beq\label{horodef1}
S^-_\infty =  \partial \bigg( \bigcup_k P_k^- \bigg) = \partial \bigg( \bigcup_k I^-(S^-_k) \bigg) \,.
\eeq
\item[(2)] If $\{F_k^-\}$ is an increasing sequence,  we obtain the past achronal limit,
\beq \label{horodef2}
S^-_\infty = \partial \bigg( \bigcup_k F_k^- \bigg) = \partial \bigg( \bigcap_k J^-(S^-_k) \bigg)\,.
\eeq
\een
In either case, (if nonempty), we refer to $S^-_\infty = \lim S^-_k$ as the (generalized) \emph{past horosphere} associated to the sequence of \emph{prehorospheres}, $\{S^-_k\}$. \emph{Future horospheres}, $S^+_\infty$, are constructed time-dually, namely, as (past or future) achronal limits of future spheres, $\{S^+_k\}$. }  
\end{Def}

We observe that, as they are achronal boundaries by construction, {\it horospheres (past or future) are edgeless achronal $C^0$ hypersurfaces.}

\medskip

We  present some further properties of (generalized)  past horospheres.   

\begin{prop}\label{pastcs}
A past horosphere $S^-_{\infty}$ that is future bounded (i.e., $S^-_{\infty} \subset J^-(S)$ for some Cauchy surface $S$)  is a  past Cauchy surface, $H^-(S) = \emptyset$. 
\end{prop}

\proof This follows immediately from Lemmas \ref{PCCiffPCS} and \ref{PCCiffFB}.\qed

\medskip 
The following result is essential to later geometric applications.

\begin{thm}\label{rays}
Suppose  $S^-_{\infty}$ is a past horosphere.  Then $S^-_{\infty}$ admits a future timelike or null    $S^-_{\infty}$-ray from each of its points.  Moreover if $S^-_{\infty}$ is future bounded then  each $S^-_{\infty}$-ray is timelike.   In this case $S^-_\infty$ is also acausal. \end{thm}

Theorem \ref{rays} is primarily a consequence of the following lemma. 

\begin{lem}[The $S_k$-Segment Lemma]\label{seglem}
Let $S_k$ be a sequence of subsets with limit set  $S$,  whereby each $s \in S$ is the limit of a sequence $s_k \in S_k$. Suppose that for each $k$, there is a maximal future $S_k$-segment, $\alpha_k$, of (Lorentzian) length $l_k$ from a point $x_k \in S_k$, with $l_k \to \infty$. If $x_k \to x \in S$, then there exists a future $S$-ray from $x$.
\end{lem}

\proof Fix a complete Riemannian metric $h$ on $M$ and, for each $k$, let $\alpha_k : [0,T_k] \to M$ be a maximal future $S_k$-segment of Lorentzian length $l_k$ from $\alpha_k(0) = x_k \in S_k$, parameterized with respect to $h$-arc length. As $M$ is taken to be global hyperbolic, $d$ is continuous. This implies $T_k \to \infty$. To see why, observe that for large $k$, the (convergent) initial points $x_k = \a_k(0)$ are contained in a compact $h$-ball around $x_\infty$. If some subsequence $T_{k_j}$ were bounded, the endpoints $\alpha_{k_j}(T_{k_j})$ would also be contained in a compact $h$-ball around $x_\infty$. Then by continuity, $d$ being bounded on the product of such balls, we would have $l_{k_j} = d(x_{k_j}, \a_{k_j}(T_{k_j})) \le C$, contradicting $l_k \to \infty$. 

Extending the $\a_k$'s arbitrarily to the future and applying the limit curve lemma, there is a subsequence $\a_{k_j}: [0, \infty) \to M$ and a future-inextendible $C^0$ causal limit curve $\a : [0,\infty) \to M$  from $\alpha(0) = x \in S$, such that $\{\a_{k_j}\}$ converges to $\a$ uniformly with respect to $h$ on compact parameter intervals.

Fix $y \in S$.  By assumption there exists a sequence $y_k \in S_k$,  such that 
$y_k \to y$.  For any $T > 0$, $\a_k(T)$ is defined for large $k$, and by the upper semi-continuity of arc length and (lower semi-)continuity of $d$, we have,
\begin{align*}
L(\a |_{[0,T]}) &\ge \limsup L(\a_{k_j} |_{[0,T]})  \\
& \ge  \liminf d(S_{k_j}, \a_{k_j}(T))\\
& \ge \liminf d(y_{k_j}, \a_{k_j}(T))  \\
& \ge d(y, \a(T))   \,.
\end{align*}

Since $y \in S$ is arbitrary, we conclude that $L(\a |_{[0,T]}) = d(S, \a(T))$ for all $T > 0$, and thus 
$\a$ is an $S$-ray.\qed

\proof[Proof of Theorem \ref{rays}.] By Proposition \ref{sequential}, each point 
$x \in S^-_{\infty}$
is a sequential limit of points $x_k \in S^-_k =  S^-_{r_k}(C_k)$.  By Lemma \ref{maxseg}, 
there is a maximal future $S^-_k$-segment of length $r_k$ based at $x_k$.  Thus, we may apply Lemma \ref{seglem} to conclude that there is a future $S^-_{\infty}$-ray from each point of $S^-_{\infty}$. 
Suppose now that $S^-_{\infty}$ is future bounded by a Cauchy surface $S$, i.e., $S^-_\8 \subset J^-(S)$. Letting $\eta$ be any future $S^-_\8$-ray, then $\eta$ must meet and enter the timelike future of $S$. In particular, $\eta$ must leave $S^-_\infty$ and hence by Lemma \ref{nullray}, $\eta$ is timelike.
That $S_\infty^-$ is acausal in this case follows by corner cutting; let $\eta : [0, \infty) \to M$ be a timelike future $S_\infty^-$-ray from $x \in S_\infty^-$. Then, by definition, each initial segment $\eta: [0,T] \to M$ is a maximal $S_\infty^-$-segment, i.e., $\eta|_{[0,T]}$ is the longest curve from $S_\infty^-$ to $\eta(T)$. If there were a point $y \in S_\infty^- \cap J^-(x)-\{x\}$, (then by cutting the corner near $x$) one could produce a longer curve from $y$, and hence from $S_\infty^-$, to $\eta(T)$.
\qed

\subsection{Cauchy and ray horospheres}

In this section we construct two important concrete classes of horospheres. The \emph{ray horosphere} is built as a limit of point spheres with centers taken along a ray, and mimics the standard Busemann level set construction (see further discussion of this below).  The \emph{Cauchy horosphere} is built instead from a compact Cauchy surface, 
$S$, and its sequence of future Cauchy spheres.

\subsubsection{Ray horospheres.} Let $\gamma : [0, \infty) \to M$ be a future complete unit speed timelike ray.
Then the sequence of {\it ray prehorospheres} $S^-_k = S^-_k(\gamma(k))$ satisfies $S^-_k \subset J^-(S^-_{k+1})$ for all $k$. To see this, let $a \in S^-_k = \{x : d(x, \gamma(k)) = k\}$.  By the reverse triangle inequality,
$$
d(a, \g(k+1)) \ge d(a, \g(k)) + d(\g(k),\g(k+1)) = k + 1  \,,
$$
from which it follows that $a \in J^-(S^-_{k+1})$. Thus, 
by Remark \ref{pastproper}, the sequence $P^-_k = I^-(S^-_k)$ is increasing, and we are led to make the following definition.

\begin{Def}[Ray Horosphere] \label{defrayhoro} 
{\rm Let $\g : [0, \infty) \to M$ be a future complete, unit speed timelike geodesic ray. Then the sequence of \emph{ray prehorospheres}, $\{S^-_k\} := \{S^-_k(\g(k))\}$ is monotonic, with increasing pasts $\{P^-_k\} = \{I^-(S^-_k)\}$, and we define the \emph{ray horosphere} associated to $\g$ to be the future achronal limit,
\beq
S^-_\infty(\gamma) = \partial \bigg( \bigcup_k P^-_k \bigg)  =
 \partial \bigg( \bigcup_k I^-(S^-_k) \bigg)  \,.
 \eeq
 }
\end{Def}

\smallskip
Note that $S^-_\infty(\gamma)$ is nonempty:  Since $\g(0) \in S^-_k$ for all $k$, it follows from 
Proposition~\ref{sequential} that $\g(0) \in S^-_\infty(\gamma)$. Future ray horospheres $S_{\infty}^+(\g)$ are defined in a time-dual manner.

Applying Proposition \ref{pastcs} and Theorem \ref{rays} one has the following.

\begin{lem}\label{rayhorolem}
$S^-_\infty(\gamma)$ is an edgeless achronal $C^0$ hypersurface which admits a future 
$S^-_\infty(\g)$-ray from each of its points.  If $S^-_\infty(\gamma)$ is future bounded by a Cauchy surface, then each of these rays is timelike and $S^-_\infty(\gamma)$ is an acausal past Cauchy surface.  In general,
 $\g$ is itself an $S^-_\infty(\gamma)$-ray.
\end{lem}

There is a basic circumstance under which ray horospheres are future bounded.

\begin{lem}\label{rayhorolem2}
Let $\gamma$ be a future complete timelike $S$-ray, for some Cauchy surface $S$, and let $S^-_\infty = S^-_\infty(\gamma)$. Then:
\ben
\item[(1)] $S^-_k  \subset  J^-(S)$, for all $k$, and hence $S^-_\infty \subset J^-(S)$.
\item[(2)] $S^-_\infty \subset I^-(\gamma)$ and $S^-_\infty$ admits a timelike future $S^-_\infty$-ray from each point.
\een
\end{lem}

\proof
To see (1), note that, parameterizing by arc length, the fact that $\gamma$ is an $S$-ray means that $d(x,\gamma(k)) \le d(\gamma(0), \gamma(k)) = k$, for all $x \in S$. Thus, for any $y \in I^+(S)$, we have $d(y,\gamma(k)) < k$, so $S^-_k = S^-_k(\gamma(k)) \subset J^-(S)$. Being the achronal limit of sets contained in the closed set $J^-(S)$, we have also $S^-_\infty \subset J^-(S)$. To see (2), note that each pre-horosphere $S^-_k$ is a past sphere from a point on $\gamma$ and is thus contained in $I^-(\gamma)$, from which it follows that $S^-_\infty \subset \overline{I^-(\gamma)}$. By Theorem \ref{rays}, $S^-_\infty$ admits a future ray from each point. In the case of a ray horosphere, each such ray is realized as a limit curve of a sequence of maximal $S^-_k$-segments which are contained entirely in $I^-(\g)$. Hence, each $S^-_\infty$-ray  is contained in $\overline{I^-(\g)}$. As $S^-_\infty$ is future bounded, all $S^-_\infty$-rays are timelike. Suppose $x \in S^-_\infty \cap \d I^-(\g)$. The future $S^-_\infty$-ray from $x$ is contained in $\overline{I^-(\g)}$, hence must remain in $\d I^-(\g)$. But as this ray is timelike, this is impossible. Hence $S^-_\infty \subset I^-(\g)$. \qed

\medskip
\noindent
{\it Remark on Busemann functions.} Given a future complete unit speed  timelike geodesic ray, $\g : [0, \8) \to M$, in a globally hyperbolic spacetime 
$M$, the associated Busemann function 
$b = b_\g$ is defined as
$$
b(x) = \lim_{k \to \8} \big[k - d(x,\g(k))\big]   \,.
$$ 
 
\noindent
Contrary to the Riemannian case, although the  pre-Busemann functions  $b_k(x) = k - d(x, \g(k))$ are continuous everywhere, the limit above in general is not (cf., \cite{Beemetal}).  In particular, while the  $0$-level sets of the pre-Busemann functions $\{b_k = 0\} =  S^-_k(\g(k))$ are well-behaved past spheres, little is known  about the Busemann level set \{$b = 0\}$ without imposing additional assumptions.  On regions where the so-called timelike co-ray condition holds, the pre-Busemann functions $b_k$ converge  uniformly on compact subsets to $b$, and hence on such regions $b$ is continuous, cf., \cite{Beemetal, Esch, GH}.  In particular, if one assumes that the timelike co-ray holds on $I^-(\g)$, then from the uniform convergence of $b_k$ to $b$ on $I^-(\g)$ one can show that the Busemann level set 
$\{b = 0\}$ and the ray-horosphere $S^-_{\8}(\g)$ agree. 
However, apart from special situations (\cite{Beemetal, EschGal}), the timelike co-ray condition is known to hold in general only in a neighborhood of $\g((0,\8))$ \cite{Esch}.  The philosophy of the present paper is to dispense with Lorentzian Busemann functions and their analytic difficulties, and to focus directly on the convergence properties of sequences of spheres. This approach is similar in spirit to the classical treament of horospheres in hyperbolic geometry.  From the approach taken here, regularity of a limit of spheres is a consequence of the causality of Lorentzian manifolds, specifically through the properties of achronal boundaries.

\subsubsection{Cauchy horospheres}  

For this construction, we begin with a compact Cauchy surface $S$ in a future timelike geodesically complete spacetime $M$.  The idea is to replace the sequence of center points, $\{\g(k)\}$, in the construction of the ray horosphere, with the sequence of future Cauchy spheres $\{S^+_k(S)\}$, then similarly take a limit of past spheres from this sequence.  
Note, by Lemma \ref{Cauchyspheres}, each $S^+_k(S)$ is a compact Cauchy surface, and, in particular, is past causally complete.  

\smallskip
We define the sequence of \emph{Cauchy prehorospheres} by
$$
\tlS_k :=S^-_k(S^+_k(S))  \,.
$$
\noindent
Again, each $\tlS_k$ is a past proper achronal boundary, with corresponding past and future sets, $\tlP_k = I^-(\tlS_k)$ and $\tlF_k$.

Like the ray prehorospheres, the sequence of Cauchy prehorospheres is monotonic, but  in the opposite direction.
To see this, let $x \in \tlS_{k+1}$. Hence, $d(x, S^+_{k+1}(S)) = k+1$. Let $\a$ be any future timelike curve from $x$ to $S^+_{k+1}(S)$ which realizes this distance. Then there is a unique point $x_k := \a \cap S^+_k(S)$. Let $\a^-_k$ be the portion of $\a$ before $x_k$ and $\a^+_k$ the portion after. By (the time-dual of) Lemma \ref{equidist}, $d(S^+_k(S), S^+_{k+1}(S)) = 1$, and hence $L(\a^+_k) \le 1$. Thus, we have, 
$$
L(\a^-_k) = L(\a) - L(\a^+_k) \ge (k+1) - 1 = k
$$

\noindent
Hence, $d(x, S^+_k(S)) \ge L(\a^-_k) \ge k$, which implies $x \in J^-(\tlS_k)$.
 Since $x$ was arbitrary, this shows $\tlS_{k+1} \subset J^-(\tlS_k)$. Hence by Remark \ref{pastproper},  $\{\tlP_k\} = \{I^-(\tlS_k)\}$ is decreasing, or, equivalently  $\{\tlF_k\}$ increasing, and thus, we have a well-defined achronal limit, 
$S^-_\infty(S) = \lim \tlS_k$. 
This leads, in summary,  to the following definition.

\smallskip
\begin{Def} [Cauchy Horosphere] \label{defCauchyhoro} 
{\rm Suppose $M$ is future timelike geodesically complete and admits a compact Cauchy surface $S$. Then the sequence of \emph{past Cauchy prehorospheres}, $\{\tlS_k\} := \{S^-_k(S^+_k(S))\}$ is monotonic, with decreasing past sets $\{\tlP_k\} = \{I^-(\tlS_k)\}$, or equivalently, increasing future sets $\{\tlF_k\}$, and we define the \emph{past Cauchy horosphere} associated to $S$ to be the past achronal limit,
\beq\label{chorodef}
S^-_\infty(S) := \d \bigg( \bigcup_k \tlF_k \bigg) = \d \bigg( \bigcap_k J^-(\tlS_k) \bigg)  \,.
\eeq
}
\end{Def}

\smallskip
Since $S$ is a compact Cauchy surface, by standard techniques one can construct a future timelike $S$-ray $\g$ (see, e.g., \cite[Lemma 4]{EschGal}).
Then we have $\g(0) \in \tlS_k$ for all $k$ and hence $\g(0) \in S^-_\infty(S)$.  In particular, 
$S^-_\infty(S)$ is always nonempty.

\smallskip
We consider some basic properties of $S^-_\infty(S)$.

\smallskip
\begin{lem}\label{horoprops} Let $S^-_\infty = S^-_\infty(S)$.
\ben
\item[(1)]  $S^-_\infty$ is future bounded by $S$. 
\item[(2)]  $S^-_\infty$ is an acausal past Cauchy surface which admits a timelike future $S^-_\infty$-ray from each point.
\item[(3)]  If $\gamma$ is a future $S$-ray, then $S^-_k(\g(k)) \subset J^-(\widetilde S_k)$ and   $S^-_\infty(\gamma) \subset J^-( S^-_\infty(S))$.
\een
\end{lem}

\proof  It is straightforward from the definitions that $\widetilde{S}_k \subset J^-(S)$ for all $k$. Hence, (1) follows from Proposition \ref{sequential}. 
(2) then follows from (1), Proposition \ref{pastcs} and Theorem~\ref{rays}. Finally, to see (3), let $\gamma$ be a future $S$-ray, which must be timelike. Parameterizing $\gamma$ by arc length, we see that $\gamma(k) \in S^+_k(S)$. One then easily checks that the past point sphere $S^-_k(\gamma(k))$ from $\gamma(k)$ lies to the past of the past Cauchy prehorosphere $\widetilde{S}_k = S^-_k(S^+_k(S))$.  (If $a \in S^-_k(\gamma(k))$, then, since $\gamma(k) \in S^+_k(S)$, the distance from $a$ to $S^+_k(S)$ is already least $k$.) Thus, the prehorospheres  satisfy,  $S^-_k(\gamma(k)) \subset J^-(\widetilde{S}_k)$ for all $k$.
The horosphere relation in (3) follows from this and Proposition \ref{sequential}.\qed 

\medskip
\noindent{\it Remark.}  Suppose now that $M$ is past, as well as future, timelike geodesically complete.  Then by Lemma \ref{Cauchyspheres}, each past Cauchy sphere  $\widetilde{S}_k$ is a compact Cauchy surface.  In particular each $\widetilde{S}_k$ is future proper, 
$ \widetilde F_k  = I^+(\widetilde{S}_k)$, and hence from Equation \eqref{chorodef},
\beq
S^-_\infty(S) = \partial \bigg( \bigcup_k  I^+(\widetilde{S}_k) \bigg) \,.
\eeq
By Proposition \ref{sequential}, $S^-_\infty(S)$ is, in this case, the sequential limit of compact
Cauchy surfaces.  However, in general $S^-_\infty(S)$ may itself fail  to be Cauchy, or, equivalently, fail to be compact.  Some criteria for compactness of $S^-_\infty(S)$ are considered in Section~\ref{horoapps}.

In the next section we establish some convexity and rigidity properties of generalized horospheres and Cauchy horospheres in particular.

\section{Convexity and Rigidity}
\label{CR}

\subsection{Weak mean curvature inequalities}

Motivated by several earlier works, in \cite{Esch89} Eschenburg introduced  the notion of mean curvature  inequalities {\it in the support sense} for rough hypersurfaces in Riemannian manifolds.  
In \cite{AGH}  Andersson, Howard and the first author considered the related situation for rough spacelike hypersurfaces in Lorentzian manifolds and proved a geometric maximum principle for such hypersurfaces.  Before stating this result we recall some basic definitions.

To set sign conventions, consider  a smooth spacelike hypersurface $\S$ in a spacetime $M$, with induced metric $h$ and second fundamental form $K$.  For $X, Y \in T_p\S$, 
$K(X,Y) = h(\D_X u, Y)$, where $u$ is the future pointing timelike unit vector field orthogonal to 
$\S$.  Then $H =$ the mean curvature of $\S = {\rm tr}_{\S} K = \div_{\S} u$. 

We adopt  the weak version of spacelike hypersurfaces used in \cite{EschGal, GH,  AGH}; see also \cite[Chapter 14]{BEE}. A  {\it $C^0$-spacelike hypersurface}  in $M$ is a subset 
$S \subset M$ that is locally acausal and edgeless, i.e.,  for each $p \in S$, there is a neighborhood $U$ of $p$ in $M$ so that $S \cap U$ is acausal and edgeless in $U$.
A $C^0$-spacelike hypersurface is necessarily an embedded $C^0$ hypersurface in $M$, as follows from \cite[Proposition 5.8]{Penrose}.  Note also that future bounded past horospheres 
$S^-_\infty$, being (globally) acausal, are $C^0$-spacelike hypersurfaces.

Let $S$ and $S'$ be  $C^0$-spacelike hypersurfaces meeting at a point $q \in S \cap S'$. We say that {\it $S'$ is locally to the future of $S$ near $q$} if, for some neighborhood $U$ of $q$ in which $S$ is acausal and edgeless, $S' \cap U \subset J^+(S, U)$. In this case we also call $S'$ a {\it future support hypersurface for $S$ at $q$}.

We say that a $C^0$-spacelike hypersurface $S$ has mean curvature $\le a$ {\it in the support sense} if for all $q \in S$ and $\epsilon > 0$, there is a smooth (at least $C^2$) future support spacelike hypersurface $S_{q,\epsilon}$ to $S$ at $q$ with mean curvature $H_{q, \epsilon}$ satisfying 
\beq\label{mean}
H_{q, \epsilon}(q) \le a + \epsilon  \,.
\eeq
Time-dually, we have the notion of a  past support hypersurface for $S$ at $q\in S$, and we may speak of a $C^0$-spacelike hypersurface $S$ having mean curvature $\ge a$ in the support sense (by requiring the reverse of inequality \eqref{mean}, $H_{q, \epsilon}(q) \ge a - \epsilon$).

\begin{thm}[\cite{AGH}] \label{AGH}
Let $S_1$ and $S_2$ be  $C^0$ spacelike hypersurfaces such that, for some constant $a$, we have,
\ben
\item[(1)] $S_2$ is locally to the future of $S_1$ near $q \in S_1 \cap S_2$.
\item[(2)] $S_2$ has mean curvature $H_2 \le a$ in the support sense.
\item[(3)] $S_1$ has mean curvature $H_1 \ge a$ in the support sense (with one-sided Hessian
bounds\footnote{See the appendix for a precise statement of this technical condition.}\!). 
\een
Then there is a neighborhood $U$ of $q$, such that $S_1 \cap U = S_2 \cap U$ and this intersection is a smooth spacelike hypersurface with mean curvature $H = a$.
\end{thm}

\subsection{A splitting result for generalized horospheres}

We begin by establishing a fundamental property of generalized horospheres.

\begin{thm}\label{convex}
Suppose $M$ is a future timelike geodesically complete spacetime satisfying the timelike convergence condition,
${\rm Ric}(X,X) \ge 0$ for all timelike vectors $X$.  If $S^-_\infty$ is a past horosphere such that every future $S^-_\infty$-ray is timelike (for example, if $S^-_\infty$ is future bounded), then $S^-_\infty$ has mean curvature $\ge 0$ in the support sense  (with one-sided Hessian bounds). 
\end{thm}

\proof Let $x \in S^-_\infty$. By Theorem \ref{rays}, we have a future timelike $S^-_\infty$-ray, $\gamma$, from $x$. By the completeness assumption, $\g$ is future complete, and, parameterizing with respect to arc length, we have $\g : [0, \infty) \to M$. Since $\g|_{[0,r]}$ is a maximal $S_{\infty}^-$-segment, one sees that the past distance sphere $S^-_r(\g(r))$  is a past support surface for $S^-_\infty$ at $x$.  
Since there are no cut points to $\g(r)$ along $\g|_{[0,r]}$, the distance function $x \to d(x,\g(r))$ is smooth in a neighborhood of  $\g([0,r))$, and in particular, the past sphere $S^-_r(\g(r))$ is smooth near $x$.  Then, using the curvature assumption, basic comparison theory (see e.g., \cite{Esch87}) implies that  $S^-_r(\g(r))$ has mean curvature $\ge - \frac{n}{r}$ at $x$, (where $M = M^{n+1}$). Since $r$ can be taken arbitrarily large, and since $x$ was arbitrary, we conclude  that $S^-_\infty$ has mean curvature $\ge 0$ in the support sense.  
The part about one-sided Hessian bounds follows from \cite[Proposition 3.5]{AGH} and the assumption that all future $ S^-_\infty$-rays are timelike; see the appendix for a more detailed discussion of this point.\qed

\medskip
The following proposition generalizes to support mean curvature inequalities Theorem~C in \cite{GalJGP}. 

\begin{prop}\label{hyprigid}
 Let $M$ be a globally hyperbolic and future timelike geodesically complete spacetime satisfying the timelike convergence condition. Let $S$ be a connected, acausal, future causally complete, $C^0$-spacelike hypersurface with mean curvature $\le 0$ in the support sense. If $S$ admits a future $S$-ray, then $S$ is a smooth, maximal, geodesically complete spacelike hypersurface, and the causal future of $S$ splits; i.e., $(J^+(S),g)$ is isometric, via the normal exponential map, to $( [0, \infty) \times S, -dt^2 \oplus h)$, where $h$ is the induced metric on~$S$.
\end{prop} 

\proof  By the time-dual of Lemma \ref{PCCiffPCS}, $S$  is a future Cauchy surface, $H^+(S) = \emptyset$.
Hence, $J^+(S) \subset D(S)$ and by restricting to the spacetime $D(S)$, we may assume without loss of generality that $S$ is a Cauchy surface for $M$.

Fix a future $S$-ray $\gamma$, which must be timelike, and hence future complete.   Let $S^-_\infty = S^-_\infty(\gamma)$ be its associated past ray horosphere. By Lemmas \ref{rayhorolem} and \ref{rayhorolem2}, $S^-_\infty$ is acausal, with $S^-_\infty \subset J^-(S)$, and all future $S^-_\infty$-rays are timelike. Theorem \ref{convex} then implies that $S^-_\infty$ has mean curvature $\ge 0$ in the support sense, with one-sided Hessian bounds. Let $S^-$ be the connected component of $S^-_\infty$ which contains $\g(0)$. Then $S \cap S^-$ is non-empty and closed. Since $S$ meets $S^-$ locally to the future near any intersection point $x \in S \cap S^-$, Theorem \ref{AGH} gives that, for some open neighborhood $U$ of $x$, we have $S \cap U =  S^- \cap U$, with this overlap being smooth, spacelike, and maximal. In particular, it follows that $S \cap S^-$ is open in both $S$ and $S^-$, and hence that $S = S^-$ is smooth, spacelike, and maximal. Consequently, the timelike future $S^-_\infty$-rays from each point of $S^- = S$ are also $S$-rays. Since $S$ is smooth, these are precisely the future normal geodesics from $S$. 
The geodesic completeness of $S$ and splitting of $J^+(S)$ then follow as in Theorem C in \cite{GalJGP}.  (Alternatively, at this stage one could apply directly standard Ricatti equation techniques to the smooth spacelike hypersurfaces $S_t = \{p \in J^+(S): d(S,p) = t\}$; see \cite[Lemma 3.1]{GalJGP}). \qed

\medskip
The previous results can now be used to establish the following splitting result for generalized horospheres. 

\begin{thm}\label{horosplit}
Let $M$ be a globally hyperbolic timelike geodesically complete spacetime which satisfies the timelike convergence condition. Suppose  $S^-_\infty$ is a future bounded (generalized) past horosphere which admits a past $S^-_\infty$-ray. Then $S^-_\infty$ is a smooth spacelike geodesically complete Cauchy surface, and $M$ splits along $S^-_\infty$, i.e., $(M,g)$ is isometric to $( \bbR \times S^-_\infty , -dt^2 \oplus h)$, where $h$ is the induced metric on $S^-_\infty$.
\end{thm} 

\proof By Theorem \ref{rays}, $S^-_{\infty}$ is a  globally acausal $C^0$ spacelike hypersurface. Moreover, by 
Theorem \ref{convex}, $S^-_{\infty}$  has support mean curvature $\ge 0$  (with one-sided Hessian bounds).  Let $S^-$ be the connected component of $S^-_\infty$ from which the past $S^-_\infty$-ray emanates. Lemma~\ref{PCCiffFB} implies that $S^-$ is past causally complete. Then, by the time dual of Proposition \ref{hyprigid}, we have that $S^-$ is a smooth, acausal, maximal, geodesically complete spacelike hypersurface, such that  $(J^-(S^-), g)$ is isometric, via the normal exponential map, to $((-\infty, 0] \times S^-, -dt^2 \oplus h)$, where $h$ is the induced metric on $S^-$. It follows from Theorem \ref{rays} and the smoothness of $S^-$ that all future normal geodesics from $S^-$ are future complete timelike $S^-$-rays. Consequently, one can similarly show that the full normal exponential image $N(S^-)$ splits as $(N(S^-), g) \approx (\bbR \times S^-, -dt^2 \oplus h)$. This product structure and the geodesic completeness of $S^-$, imply that $N(S^-)$ is geodesically complete and globally hyperbolic (see e.g. \cite[Section 3.6]{BEE}). It follows that $N(S^-) = M$ and that $S^-$ is a Cauchy surface for $M$. Finally, if $x \in S^-_\infty \setminus S^-$, then $x \in I^\pm(S^-)$, which would violate the achronality of $S^-_{\infty}$. Thus, $S^- = S^-_\infty$.\qed

\medskip
As it is an edgeless achronal $C^0$ hypersurface, a compact horosphere $S^-_{\8}$ is necessarily a (compact) Cauchy surface.  As such it admits a past $S^-_{\8}$-ray and is trivially  future bounded.  Thus, we have the following corollary to Theorem \ref{horosplit}.

\begin{cor}\label{horosplitcor}
Let $M$ be a globally hyperbolic timelike geodesically complete spacetime which satisfies the timelike convergence condition. Suppose $S^-_\infty$ is a \emph{compact} past horosphere in $M$. Then $S^-_\infty$ is a smooth spacelike geodesically complete Cauchy surface, and $M$ splits along $S^-_\infty$. i.e., $(M,g)$ is isometric to $( \bbR \times S^-_\infty , -dt^2 \oplus h)$, where $h$ is the induced metric on $S^-_\infty$.
\end{cor}

\subsection{Application to Cauchy horospheres}\label{horoapps}

Let $M$ be a timelike geodesically complete spacetime with compact Cauchy surface $S$, and consider the associated Cauchy horosphere $S^-_\infty = S^-_\infty(S)$.  
We  summarize some properties of $S^-_\infty$, based on previous results.
\ben
\item[(i)] $S^-_\infty$ is the (nonempty) sequential limit of compact Cauchy surfaces.
\item[(ii)]  $S^-_\infty$ is future bounded by $S$, $S^-_\infty \subset J^-(S)$.
\item[(iii)] $S^-_\infty$ is a past Cauchy surface and admits a timelike future $S^-_\infty$-ray from each point.  
\item[(iv)] If the timelike convergence condition holds then $S^-_\infty$ has mean curvature 
$\ge 0$ in the support sense (with one-sided Hessian bounds).
\een

\smallskip 

As can be seen from Corollary \ref{horosplitcor}, compactness is a particularly consequential property for horospheres.
However, as noted earlier, although it is a limit of compact Cauchy surfaces, a Cauchy horosphere need not itself be compact in general.  Here we  present a  simple criterion for compactness via a `max-min condition' on its base Cauchy surface. 

\begin{lem}
Let $M$ be future timelike geodesically complete with compact Cauchy surface $S$.  Then $S^-_{\8}(S)$ is a compact Cauchy surface if and only if it is past bounded.
\end{lem}

\proof A Cauchy surface is trivially past bounded. Suppose conversely that $S^-_{\8}(S)$ is past bounded, i.e., $S^-_{\8}(S) \subset J^+(S'$) for some  Cauchy surface $S'$, which is necessarily compact.  Then, by (ii) above, we have $S^-_{\8}(S) \subset  J^+(S') \cap J^-(S)$.  Hence,  $S^-_{\8}(S)$ is a (closed) edgeless achronal $C^0$ hypersurface contained in compact set, and consequently must be a compact Cauchy surface.\qed

\medskip
\begin{Def} [Max-Min Condition]\label{MMdef}
 {\rm Let $M$ be future timelike geodesically complete with compact Cauchy surface $S$. For each positive integer $k$, let $S_k := S^+_k(S)$. We say the \emph{max-min condition} holds on $S$ if there is an $R > 0$, such that for all $k$,
$$
\max_{x \in S} d(x,S_k) - \min_{x \in S} d(x, S_k)  \le  R  \,.
$$
}
\end{Def}

We note that, by definition of $S_k = S^+_k(S)$, we have $\max_{x \in S} d(x,S_k) = k$. The max-min condition is easily seen to hold for any Cauchy surface in a Lorentzian warped product $(\bbR \times N, -dt^2 + f^2(t)h)$, with $f : \bbR \to (0, \8)$ and $(N,h)$ compact Riemannian. In particular, it holds for any Cauchy surface in de Sitter space.

\smallskip
\begin{lem} \label{MMlem} Let $M$ be timelike geodesically complete with compact Cauchy surface $S$. If the max-min condition holds on $S$, then $S^-_\8(S)$ is past bounded and hence is a compact Cauchy surface.
\end{lem}

\proof Suppose that $\max_{x \in S} d(x,S_k) - \min_{x \in S} d(x, S_k) = k - \min_{x \in S} d(x, S_k) \le R$, for some $R > 0$. Note that $S^-_R(S)$ is a compact Cauchy surface by Lemma \ref{Cauchyspheres}. Since  $S^-_\8(S)$ is the sequential limit of the $\widetilde{S}_k$'s, it is sufficient to show that $\widetilde{S}_k \subset J^+(S^-_R(S))$. Suppose otherwise, that there is some $x_1 \in \widetilde{S}_k$ and $x_2 \in S^-_R(S)$, with $x_1 << x_2$. By definition of $S^-_R(S)$, there is a timelike curve of length $R$ from $x_2$ to $x_3 \in S$. Then, there is a timelike curve from $x_3$ to $x_4 \in S^+_k(S)$ of length at least $\min_{x \in S}d(x,S^+_k(S))$. Concatenating these curves, we get a curve from $x_1 \in \widetilde{S}_k = S^-_k(S^+_k(S))$ to $x_4 \in S^+_k(S)$ of length strictly greater than $R + \min_{x \in S}d(x,S^+_k(S))$, and hence, $k = d(x_1, S^+_k(S)) > R + \min_{x \in S}d(x,S^+_k(S))$, a contradiction.\qed

\medskip
Corollary \ref{horosplitcor} and Lemma \ref{MMlem} combine to give the following proof of the Bartnik splitting conjecture, subject to the max-min condition.

\begin{thm}\label{maxminsplit}
Let $M$ be a timelike geodesically complete spacetime which satisfies the timelike convergence condition. If $S$ is a compact Cauchy surface on which the max-min condition holds, then the Cauchy horosphere $S^-_\infty = S^-_\infty(S)$ is a smooth compact spacelike Cauchy surface and $(M,g)$ is isometric to $( \bbR \times S^-_\infty , -dt^2 \oplus h)$, where $h$ is the induced metric on $S^-_\infty$.
\end{thm}

Working in a similar context, we recall that a splitting result was obtained in \cite{EschGal} from the following `$S$-ray condition': For some future complete timelike $S$-ray~$\gamma$, $S \subset I^-(\gamma)$. We observe here that this condition implies the max-min condition above:

\medskip
\begin{lem}\label{hierarchy}
Let $M$ be timelike geodesically complete, $S \subset M$ a compact Cauchy surface, and $S^-_\infty = S^-_\infty(S)$ its associated Cauchy horosphere. If $\gamma$ is a timelike future $S$-ray such that $S \subset I^-(\gamma)$, then the max-min condition holds on $S$.
\end{lem}

\proof Parameterize $\gamma$ with respect to arc length. Since $S \subset I^-(\gamma)$ and $S$ is compact, we have $S \subset I^-(\gamma(k_0))$ for some $k_0 \in \bbN$. Then, for any $x \in S \subset I^-(\gamma(k_0))$, and $k_0 \le k$, the reverse triangle inequality gives $d(x, \gamma(k_0)) + (k-k_0) \le d(x, \gamma(k))$, and rewriting, we get $k - d(x, \gamma(k)) \le k_0 - d(x, \gamma(k_0))$. As the right hand side is a continuous function on the compact set $S$, it is bounded above by some $0 \le R$, and we get, $k - d(x, \gamma(k)) \le R$, for all $k_0 \le k$. Since $d(x, \gamma(k)) \le d(x, S^+_k(S))$, we have $k \le d(x, S^+_k(S)) + R$. Taking the minimum over $x \in S$ gives the result. \qed

\medskip
Within the class of spacetimes considered in Lemma \ref{hierarchy} the $S$-ray condition is strictly stronger than the max-min condition.  For example,  while the max-min condition holds for any Cauchy surface $S$ in de Sitter space, the $S$-ray condition fails for all such $S$.

 \smallskip
We conclude this section with one further splitting result.  By using Theorem 3.7 in 
\cite{GalBanach}, the requirement of the existence of a past  $S^-_\infty$-ray in Theorem \ref{horosplit} (specialized to Cauchy horospheres) can be replaced by a somewhat different ray condition.

\begin{thm}\label{horosplit3} Let $M$ be a  timelike geodesically complete spacetime which satisfies the timelike convergence condition and has a compact Cauchy surface $S$.  Suppose there is a past timelike ray $\g$ emanating from a point in $I^-(S^-_{\infty}(S))$ such that the future ray horosphere 
$S^+_{\infty}(\g)$  is past bounded by some Cauchy surface $S'$.   Then $S^-_\infty(S)$ is a smooth compact Cauchy surface, and $M$ splits along  $S^-_\infty$ as in Theorem \ref{horosplit}.
\end{thm} 

\proof By previous results (or their time-duals), $\S_1 := S^+_{\infty}(\g)$ is an edgeless acausal $C^0$  hypersurface with mean curvature $\le 0$ in the support sense and  
$\S_2 := S^-_{\infty}(S)$ is a is an edgeless acausal $C^0$  hypersurface with mean curvature 
$\ge 0$ in the support sense.  Since $\S_1 \subset J^+(S')$ and 
$\S_2 \subset J^-(S)$, it follows that 
$J^+(\S_1)\cap J^-(\S_2)$ is a subset of the compact set $K = J^+(S')\cap J^-(S)$.
The equality $J^+(\S_1)\cap J^-(\S_2) = J^+(\S_1\cap K)\cap J^-(\S_2\cap K)$ is easily verified
and shows that $J^+(\S_1)\cap J^-(\S_2)$ is compact.  Since $d(\S_1, \S_2) = \delta > 0$, it now follows from  Theorem 3.7 in \cite{GalBanach}  that  $\S_1$ and $\S_2$ are smooth  compact  spacelike Cauchy  surfaces and $J^+(\S_1) \cap J^-(\S_2)$ is isometric to
a Lorentzian product ($[0, \delta] \times \S_1, -dt^2 \oplus h)$, where $h$ is the induced metric on $\S_1$.  In particular $\S_2$ is maximal (in fact totally geodesic).  Theorem \ref{horosplit3} now follows from e.g., \cite[Corollary 1]{Bart88} (or apply 
Corollary \ref{horosplitcor}).\qed

\section{Lines}
\label{lines}

In this section we present an alternative proof of Theorem \ref{horosplit}, when specialized to Cauchy horospheres, based on the Lorentzian splitting theorem (see \cite{BEE} and references therein).

\begin{thm}\label{lst}
(Lorentzian Splitting Theorem) If $(M, g)$ is a globally hyperbolic spacetime which satisfies the timelike convergence condition and admits a complete timelike line, then $(M,g)$ splits isometrically as a product, $(M,g) \approx (\bbR \times N, -dt^2 \oplus h)$,  where $(N, h)$ is a complete Riemannian manifold.
\end{thm}

Recall, a causal line in spacetime is an inextendible causal geodesic with the property that every segment maximizes the Lorentzian distance between its endpoints.

The approach taken here rests on the following basic property of generalized horospheres.

\begin{prop}\label{horoline}
Let $S^-_\infty$ be a generalized past horosphere in a globally hyperbolic spacetime $M$. Then any past timelike $S^-_\infty$-ray $\gamma$ extends to a timelike line.
\end{prop}

\proof 
The proof is a direct consequence of Theorem \ref{rays}. $S^-_\infty$ has a future $S^-_\infty$-ray $\eta$ extending from $\gamma(0)$.   By Proposition \ref{decomp}, any causal curve $\a : [a,b] \to M$ from $x = \g(t)$ to $y = \eta(s)$ must meet $S^-_\infty$ at some point $p = \a(c)$, say. But then, because $\eta$ and  $\gamma$ are $S^-_\infty$-rays, we have that
$$L(\alpha) = L(\alpha |_{[a,c]}) + L(\alpha|_{[c, b]}) \le d(x, S^-_\infty) + d(S^-_\infty, y) = 
 L(\g|_{[0,t]}) + L(\eta|_{[0, s]})
$$

It follows that joining $\eta$ and $\gamma$ produces a causal line, which is necessarily  timelike since $\gamma$ is.\qed

\smallskip
We shall also need the following lemma, which shows, in particular,  that, for the splitting in 
Theorem~\ref{linesplit} below, compactness of the Cauchy horosphere  $S^-_\infty(S)$  is both necessary and sufficient.

\begin{lem}\label{compacthoro}
Suppose  $(M, g)$ is isometric to the Lorentzian product  $(\bbR \times N, -dt^2 \oplus h)$, for some compact, connected Riemannian manifold $(N,h)$. For any Cauchy surface $S$ in $M$, the associated Cauchy horosphere $S^-_{\infty} = S^-_\infty(S)$ is a compact Cauchy surface.
\end{lem}

\proof  Since $N$ is compact, $(M,g)$ is globally hyperbolic with compact Cauchy surfaces.   Hence, $S$ is necessarily compact and there is some slice $N' : = \{a\}\times N$  which lies to the past of $S$,  $N' \subset J^-(S)$.  From the product structure of $(M,g)$ and the fact that $N'$ is a slice, we have  that  $\widetilde{N'}_k = N'$.  It follows  that  $\widetilde{S}_k \subset J^+(N') \cap J^-(S)$,
and hence, by Proposition \ref{sequential}, that $S^-_\infty  \subset J^+(N') \cap J^-(S)$.  As an edgeless acausal $C^0$  hypersurface contained in a compact set, $S^-_\infty$ must itself be a compact Cauchy surface.\qed

\begin{thm} \label{linesplit} Let $M$ be a future timelike geodesically complete spacetime, satisfying $\mathrm{Ric}(X,X) \ge 0$ for all timelike vectors $X$. Suppose $M$ admits a compact Cauchy surface $S$ and that its associated Cauchy horosphere $S^-_\8(S)$ admits a complete past $S^-_\8(S)$-ray. Then $S^-_\8(S)$ is a smooth, compact spacelike Cauchy surface, along which $M$ splits.
\end{thm}

\proof
By assumption, we have a past $S^-_\infty(S)$-ray $\gamma$. Since $S^-_\infty(S)$ is acausal, this ray must be timelike.  Joining this with any future $S^-_\infty(S)$-ray emanating from the base point of 
$\gamma$ produces a timelike line, as in Proposition \ref{horoline}.  Then, by Theorem~\ref{lst},  
$M$ splits isometrically as a product, $(M,g) \approx (\bbR \times N, -dt^2 \oplus h)$,  where $(N, h)$ is a complete Riemannian manifold.  Since the Cauchy surfaces of $M$ are compact, $N$ must be compact.  Hence, by Lemma \ref{compacthoro},  $S^-_\infty(S)$ is a compact Cauchy surface.  It remains to observe that $S^-_\infty(S)$ is a $t$-slice.
Along $S^-_\infty(S)$, the time coordinate $t$ achieves  a maximum value, $t = b$, say, at some point $p \in  S^-_\infty(S)$.
The slice $N_b = \{b\} \times N$ has zero mean curvature and, by Theorem \ref{convex},  
$S^-_\infty(S)$ has mean curvature $\ge 0$ in the support sense.  The geometric  maximum principle, Theorem \ref{AGH}, then implies that  $S^-_\infty(S)$ and $N_b$ agree near $p$.  In fact, by a straightforward continuation argument,  one has, $S^-_\infty(S) = N_b$.  Theorem \ref{linesplit} follows.\qed

\section{The case of a positive cosmological constant}
\label{lambda}

\subsection{Rigid singularity result for asymptotically dS spacetimes}

In this section we consider spacetimes $(M^{n+1},g)$ which obey the 
Einstein equation,
\beq\label{einstein}
R_{ij} -\frac12Rg_{ij} +\Lambda g_{ij} = 8\pi T_{ij} \,,
\eeq
with  positive cosmological constant $\Lambda$, where the energy-momentum tensor
$T_{ij}$  is assumed to satisfy  the strong energy condition, 
\beq\label{SEC}
(T_{ij} - \frac1{n-1}Tg_{ij})X^iX^j \ge 0 \, 
\eeq
for all timelike vectors $X$, where $T = T_i{}^i$.

Setting $\Lambda = n(n-1)/2{\ell^2}$, the strong energy condition~(\ref{SEC}) is
equivalent to,
${\rm Ric}\,(X,X) = R_{ij}X^iX^j \ge -\frac{n}{\ell^2}$
for all unit timelike vectors $X$. By rescaling the metric, we may set $\ell = 1$ so that $\Lambda = n(n-1)/2$ and the spacetime Ricci tensor satisfies,
\beq\label{EC}
\ric (X,X) \ge -n  \quad \text{for all unit  timelike vectors } X.
\eeq

In \cite{AndGal, Galbeem} results are presented which establish connections between the geometry and topology of spacelike conformal infinity $\scri^+$ and the occurrence of past singularities in future asymptotically de Sitter (dS) spacetimes.  A related result is obtained here, which does not require the explicit introduction of conformal infinity.  

As described in \cite{AndGal, Galbeem},  there is a connection between the occurrence of past singularities  in an asymptotically dS spacetime and the scalar curvature of its Cauchy surfaces.
Let $\S$ be a smooth compact Cauchy surface in $(M^{n+1},g)$ which has positive mean curvature,  $H > 0$.  If $(M,g)$ satisfies the dominant energy condition, the Hamiltonian constraint implies,
\beq 
H^2  \ge 2\Lambda + |K|^2 - S_\Sigma \,,
\eeq
where $S_\Sigma$ and $K$ are the scalar curvature and second fundamental form of $\S$, respectively.   Using 
$\Lambda = n(n-1)/2$, and  $|K|^2 \ge H^2/n$ (by the Cauchy-Schwartz inequality) in the above implies,
\beq
H \ge \sqrt{n^2 -\frac{n}{n-1} S_\Sigma} \,.
\eeq

Thus, if  $\S$ has negative scalar curvature then $H > n$.  Assuming \eqref{EC} holds, a straightforward generalization \cite{AndGal, Borde} of the Hawking singularity theorem  then implies that all timelike geodesics are past incomplete.  If $\S$ is only assumed  to have nonpositive scalar curvature then one obtains a rigid singularity result (\cite[Proposition 3.4]{AndGal}): Either the normal geodesics to $\S$ are past incomplete, or else the metric assumes a particular warped product structure, as exemplified by the `de Sitter cusp' discussed in  \cite{Galbeem}.  In either case, $M$ is past timelike geodesically incomplete. 

Consider, on the other hand, standard de Sitter space, 
\beq\label{deS}
M = \Bbb R \times S^n, \qquad ds^2 = -dt^2 + \cosh^2t \,d\Omega^2  \,,
\eeq
which is timelike geodesically complete.
The constant $t$-slices $S_t = \{t\} \times S^n$ are round spheres and hence have positive scalar curvature.  The mean curvature $H_t$ of  the slice $S_t$ is given by,
\beq\label{dSmean}
 H_t = n \dfrac{(\cosh t)^\prime}{\cosh t} = n \tanh t  \,.
\eeq 
Thus, $H_t < n$ but approaches $n$ rapidly as $t \to \infty$; a brief computation shows,
\beq\label{bigO}
H_t = n + O(e^{-2t})  \,.
\eeq

The following theorem (which was motivated, in part, by the Riemannian result, Theorem 3 in \cite{zero}) shows in effect that if the mean curvature converges any more rapidly to the value $n$ 
then there will be past singularities. 

\begin{thm} \label{dSrigid}
 Let $(M^{n+1},g)$ be a future timelike geodesically complete spacetime satisfying the energy condition \eqref{EC}.
 Let $S$ be a compact Cauchy surface such that the future Cauchy spheres $S^+_k(S)$ have support mean curvature $\ge a_k$, where, letting $n_k := \min\{a_k, n\}$, we have
\beq\label{littleo}
n_k = n + o(e^{-2k})  \,.
\eeq
Let $S^-_\infty = S^-_\infty(S)$ be the past Cauchy horosphere associated to $S$, and suppose that 
$S^-_\infty$ admits a past $S^-_\infty$-ray~$\gamma$. Then either
\ben
\item[(1)] $S^-_\infty$ has a past incomplete timelike $S^-_\infty$-ray, or
\item[(2)]   $S^-_\infty$ is a smooth, compact spacelike Cauchy surface,
and $(M,g)$ is isometric to the warped product $(\bbR \times S^-_\infty, - dt^2 \oplus e^{2t}h)$, where $h$ is the induced metric on $S^-_\infty$.
\een
In either case, $M$ is timelike past incomplete.
\end{thm}

As the setting of Theorem \ref{dSrigid} presents some (interesting) new obstacles, a bit more work is needed before proceeding to its proof. 
The comparison techniques used to prove the splitting results in Section \ref{CR} no longer directly apply: They lead in the present setting to (weak) mean curvature inequalities for which the maximum principle (in any form)  is not relevant.  Moreover, in proving this past singularity result, we are forced to do without the assumption of (full) past completeness.

To establish Theorem \ref{dSrigid}, we will again work with horospheres, but in order to deal with the aforementioned issues, we introduce the notion of `limit mean curvature', which is adapted from an approach taken in \cite{zero}. In 
Lemma \ref{limitmax} we establish a maximum principle (of sorts) for this setting as a consequence of a key convexity result, Lemma~\ref{limconvex}, taken together with Bartnik's \cite{Bartnik.acta} solution to the Dirichlet problem for prescribed mean curvature with rough boundary data. We develop this basic framework for general achronal limits first before specializing to horospheres in Theorem \ref{dSrigid}.

\subsection{Limit mean curvature and the proof of Theorem \ref{dSrigid}}
\label{LMC}

We begin by observing that the convergence of Proposition \ref{sequential} is locally uniform.

\begin{lem} \label{uniform}Let $A_\infty$ be the (future or past) achronal limit of a sequence of achronal boundaries, $\{A_k\}$. For any neighborhood $U$ of $A_\infty$ and any compact set $K$, there is a $k_0 \in \bbN$, such that, for all $k \ge k_0$, 
$$
A_k \cap K \subset U \cap K \,.
$$
\end{lem}

\proof Otherwise, for each $j \in \bbN$, we can find $x_j \in A_{k_j} \cap K$ with $x_j \not \in U$. As $\{x_j\} \subset K$, the sequence $\{x_j\}$ has a limit point $x \in K$.  But by Proposition \ref{sequential}, we have $x \in A_\infty$, which contradicts the fact that $\{x_j\}$ never enters the neighborhood $U$.\qed

\bigskip
The following definition is adapted from \cite{zero}. We note that an achronal boundary is a $C^0$ spacelike hypersurface, (i.e., locally acausal and edgeless), if and only if it is (globally) acausal. 

\begin{Def}\label{limitmeancurv}Let $A_\infty$ be the (future or past) achronal limit of a sequence of achronal boundaries, $\{A_k\}$, each of which is acausal. We say that $A_\infty$ has \emph{limit} mean curvature $\ge a$ (resp. $\le a$) if $A_k$ has mean curvature $\ge a_k$ (resp. $\le a_k$) in the support sense, with $a_k \to a$.
\end{Def}

\smallskip
\begin{lem} \label{limithoro} Suppose $(M^{n+1}, g)$ is globally hyperbolic and satisfies \eqref{EC}. Then any future point sphere $S^+_r(p)$ has mean curvature $\le n \cdot \coth(r)$ in the support sense. Similarly, any past point sphere $S^-_r(p)$ has mean curvature $\ge -n \cdot \coth(r)$ in the support sense. Consequently, any future horosphere $S^+_\infty$ has limit mean curvature $\le n$ and any past horosphere $S^-_\infty$ has limit mean curvature $\ge - n$. 
\end{lem}

\proof Let $x \in S^+_r(p)$ and let $\a : [0, r] \to M$ be a future-directed maximal unit speed geodesic from $p = \a(0)$ to $x = \a(r)$. For $0 < \e < r$, let $\rho_\e(y) = d(\a(\e), y)$. Then $\rho_\e$ is smooth near $\a|_{(\e, r]}$. Letting $\th(t)$ be the mean curvature of the level set $\{\rho_\e = t\}$ at the point $\a(t+\e)$, then $\th = \th(t)$ satisfies the Raychaudhuri inequality:
$$
\th^\prime \le - \mathrm{Ric}(\a^\prime, \a^\prime) - \dfrac{\th^2}{n}  \le n - \dfrac{\th^2}{n}  \, .
$$
\noindent
Letting $\Theta(t) :=\th(t) / n$, we have $\Theta^\prime \le 1 - \Theta^2$. With the initial condition, 
$\lim_{t \to 0^+} \Theta(t) = \infty$, the elementary comparison solution is $\coth(t)$ (see especially 
\cite[Corollary 1.6.3]{Karcher}).  Thus, 
$$
\th(r - \e) = n \cdot \Theta(r- \e) \le n \cdot \coth(r- \e)  \,.
$$
As $\{\rho_\e = r- \e\}$ is a future support hypersurface for $S^+_r(p)$ at $\a(r)= x$, this shows that $S^+_r(p)$ has support mean curvature $\le n \cdot \coth (r-\e)$ at $x$. But since, $x \in S^+_r(p)$ and $\e > 0$ were arbitrary, $S^+_r(p)$ has support mean curvature $\le n \cdot \coth (r)$ at each point. \qed

\subsubsection{A Limit Mean Convexity Lemma}

We now establish a key convexity result which will be used to prove Lemma \ref{limitmax}. 

\begin{lem}\label{limconvex} Let $M^{n+1}$ be a globally hyperbolic spacetime such that $\mathrm{Ric}(X,X) \ge -n$ for all timelike unit vectors $X$. Let $A_\8 \subset M$ be an achronal limit with limit mean curvature $\ge n$, 
(resp. $\le n$), and suppose that $W$ is a domain in $A_\8$ with $\overline{W}$ acausal and $\overline{D(W)}$ compact. Let $\S \subset D(W)$ be a smooth, achronal spacelike hypersurface with $\mathrm{edge} \, \S = \mathrm{edge} \, W$ and mean curvature $H_\S = n$. Then $\S \subset J^+(W)$. In particular, $\S \subset J^+(A_\8)$, (resp. $\S \subset J^-(W) \subset J^-(A_\8)$). 
\end{lem}

\proof 
Suppose to the contrary that $\S$ meets $I^-(W)$. Hence, we have the (schematic) picture: 

\begin{center}
\includegraphics[width=9cm]{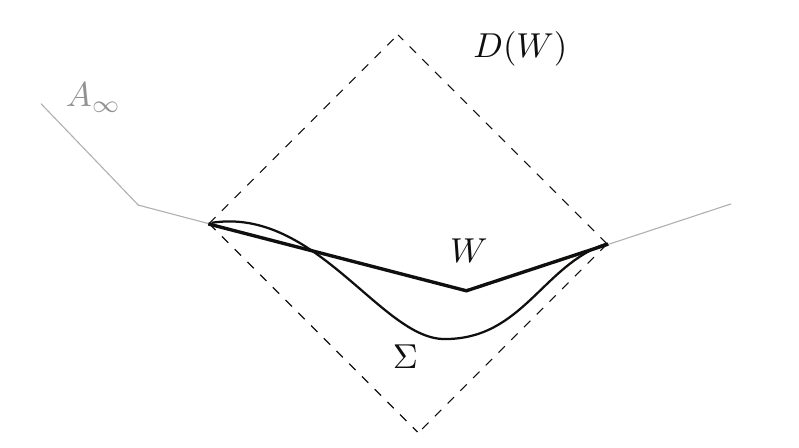}
\end{center}

The idea of the proof is as follows. We perturb (part of) $\S$ to get a smooth hypersurface with mean curvature \emph{strictly less than} $n$. That $A_\8$ has limit mean curvature $\ge n$, means $A_k$ has support mean curvature $\ge n + c_k$, with $c_k \to 0$. Then, `sliding down' a past support hypersurface for $A_k$, for large enough $k$, gives a past support hypersurface for the perturbed $\S$, with lower mean curvature bound 
arbitrarily close to $n$, producing a contradiction. (The curvature condition is used to control the mean curvature during sliding.) This will involve a bit of careful setup first.

\smallskip
Since $\S, W \subset \overline{D(W)}$, the closures $\overline{\S}$ and $\overline{W}$ are compact, and hence the distance $\ell := d(\overline{\S}, \overline{W}) \ge d(\S, W) > 0$ is realized by points $p \in \overline{\S}$ and $q \in \overline{W}$. But since $\overline{\S} = \S \cup \mathrm{edge} \, \S$ and $\overline{W} = W \cup \mathrm{edge} \, W$, and since $\mathrm{edge} \, \S = \mathrm{edge \,} W$, we must have $p \in \S$ and $q \in W$ and thus,
$$
\ell = d(p,q) = d(\S, W)    \,.
$$

Since $\overline{W}$ is acausal and compact, its `signed distance function',
$$
\delta(x) := d(\overline{W}, x) - d(x, \overline{W}),
$$
is continuous on all of $M$, and we have:

\vspace{-.5pc}
\begin{displaymath}
   \delta(x) = \left\{
     \begin{array}{ll}
       + & \; \; \; \; \; x \in I^+(\overline{W})\\
       0 & \; \; \; \; \; x \not \in I^-(\overline{W}) \cup I^+(\overline{W})\\
       - & \; \; \; \; \; x \in I^-(\overline{W})  \,.
     \end{array}
   \right.   
\end{displaymath} 

\vspace{.5pc}
\noindent
Hence, for any $a > 0$, the set $\{|\delta| < a\}$ is an open neighborhood of $\overline{W}$ (and by achronality, all of $A_\8$). Consider an exhaustion of $\S$ by smooth compact domains. Then, using the fact that $\S \cap \{ |\delta| \ge \ell/4\} = \overline{\S} \cap \{ |\delta| \ge \ell/4\}$ is compact, let $\S_0 \subset \S$ be a smooth compact domain with $\d \S_0 \subset \{|\delta| < \ell / 4\}$ and $p \in \S_0$. Hence, one still has $d(\S_0, W) = d(p,q) = \ell$. 

\begin{center}
\includegraphics[width=9cm]{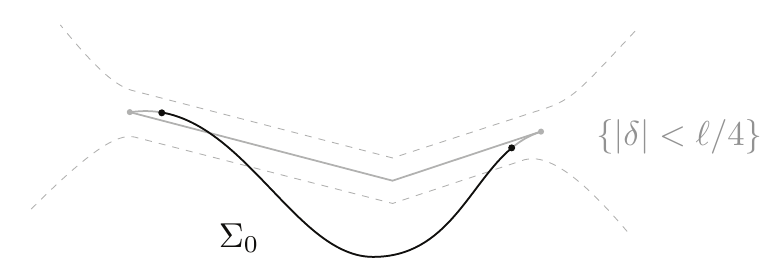}
\end{center}

For sufficiently small $f \in C^\8(\S_0)$, with $f|_{\d \S_0} = 0$, let $\mathcal{H}(f)$ denote the mean curvature of the surface $\S_f : x \to \mathrm{exp}_x f N_x$ where $N$ is the future unit normal to $\S_0$. The mean curvature operator $\mathcal{H}$ has linearization (cf. \cite{Bartnik.acta}):
$$
\mathcal{H}'(0) = \L - (\mathrm{Ric}(N,N) + |B|^2)  \,,
$$

\noindent
where $B$ denotes the second fundamental form of $\S_0$. Since, 
$\mathrm{Ric}(N,N) + |B|^2 \ge - n + \frac{H^2}{n} = 0$,
$\mathcal{H}'(0)$ is invertible. Thus, by the inverse function theorem, for sufficiently small $\e > 0$, there exists a smooth compact spacelike hypersurface $\S_\e \subset D(W)$, with $\d \S_\e = \d \S_0$ and mean curvature 
$H_{\S_\e} =  n(1-\e)$, and such that
$$
\ell_\e := d(\S_\e, \overline{W}) = d(p_\e,q_\e) \ge \dfrac{7}{8} \ell   \,,
$$
for some $p_\e \in \mathrm{int\,} \S_\e$ and $q_\e \in \overline{W}$.

\begin{center}
\includegraphics[width=9cm]{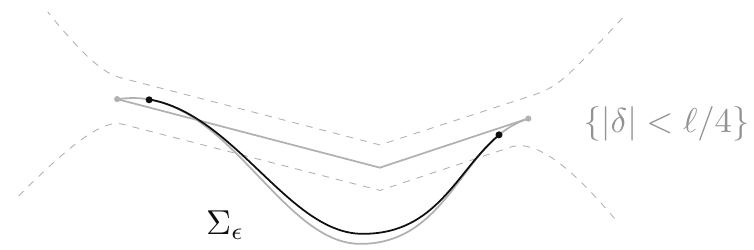}
\end{center}

Since $\S_\e$ and $\overline{D(W)}$ are compact, with $\S_\e \subset D(W)$, the set $J^+(\S_\e) \cap \d D(W)$ is compact and contained in $I^+(W)$. Hence, the signed distance function $\delta$ of $\overline{W}$ achieves a positive minimum $\delta_0 > 0$ on $J^+(\S_\e) \cap \d D(W)$. Let $\delta_1 := \min \{\delta_0, \ell /4\}$. By Lemma \ref{uniform}, we may choose $k_1$ sufficiently large so that 
$$
A_k \cap \overline{D(W)} \; \subset \; \{|\delta | < \delta_1\} \cap \overline{D(W)}, \; \; \mathrm{for \; all \; } k \ge k_1  \,.
$$
Thus, for all $k \ge k_1$, we have,
\begin{eqnarray}
  J^+(\S_\e) \cap \big(A_k \cap \d D(W)\big) & \subset & \big(J^+(\S_\e) \cap \d D(W)\big) \, \cap \, \big(A_k \cap \d D(W) \big) \nonumber\\
    & \subset &  \{\delta \ge \delta_0\}  \, \cap \, \{|\delta| < \delta_1\} \nonumber\\
    & \subset &  \{\delta \ge \delta_1\}  \, \cap \, \{|\delta| < \delta_1\} \nonumber\\
    & = &  \emptyset    \,.\nonumber 
 \end{eqnarray}
Hence for all $k \ge k_1$, 
\beq\label{canonlyseeint}
J^+(\S_\e) \cap \big(A_k \cap \overline{D(W)} \big) \subset A_k \cap D(W)\,.
\eeq

We now show that, for large $k$, the distance between the compact sets $\S_\e$ and $A_k \cap \overline{D(W)}$ remains bounded away from 0 and $\8$, and is realized by points $p_k \in \mathrm{int} \, \S_\e$ and $q_k \in A_k  \cap D(W)$. Let $\s : [0, \ell_\e] \to M$ be a future-directed maximal timelike unit-speed geodesic segment from $\s(0) \in \S_\e$ to $\s(\ell_\e) \in \overline{W}$,  realizing the distance $d(\S_\e, \overline{W}) = \ell_\e$. Since $\overline{W} \subset A_\8$,  $\s$ is a timelike curve from $\S_\e$ to $A_\8$. 

Recall,  $A_\8$ may be either a  past or a future achronal limit.  To cover both cases, extend $\s$ slightly to the future to a timelike curve, $\s : [0,L] \to M$, with $\ell_\e < L$. Then, as in Proposition \ref{sequential}, there is an integer $k_\e \ge k_1$ such that (the extended) $\s$ meets $A_k$ for all $k \ge k_\e$. Hence, for $k \ge k_\e \ge k_1$, we have $\s \cap A_k \cap \overline{D(W)} \subset \{|\delta| < \delta_1 \} \subset \{|\delta| < \ell / 4\}$, and it follows that:
$$
d(\S_\e, A_k \cap \overline{D(W)}) \ge \ell_\e - \frac{\ell}{4} \ge \frac{7\ell}{8} - \frac{\ell}{4} = \frac{5 \ell}{8}  \,.
$$
Now, for each $k \ge k_\e$, by compactness, we may find points $p_k \in \S_\e$ and $q_k \in A_k \cap \overline{D(W)}$ such that $\ell_k := d(p_k, q_k) = d(\S_\e, A_k \cap \overline{D(W)})$. But since $k_\e \ge k_1$, it follows from \eqref{canonlyseeint} that we must have $q_k \in A_k \cap D(W)$. Furthermore, since $\d \S_\e \subset \{|\delta| <  \ell/4\}$, it follows that we must have $p_k \in \mathrm{int} \, \S_\e$. Then, letting $\ell_W := d(\S_\e, \overline{D(W)})$, we have, for all $k \ge k_\e$,  
$$
\ell_k = d(p_k, q_k) = d(\S_\e, A_k \cap \overline{D(W)}) = d(\mathrm{int} \, \S_\e, A_k \cap D(W)),
$$
with,
$$
\dfrac{5 \ell}{8} \le \ell_k \le \ell_W   \,.
$$

\smallskip
Again, the idea of the last part of the proof
is to take the support hypersurfaces for $A_k$ at $q_k$, and `slide them down' to support hypersurfaces for $\S_\e$ at $p_k$. Hence, let $V_k \subset J^-(A_k) \cap D(W)$ be a (small) smooth spacelike past support hypersurface for $A_k$ at $q_k$. Since $A_\8$ has limit mean curvature $\ge n$, by choosing $k \ge k_\e$ sufficiently large, we can take $H_{V_k}(q_k) \ge n(1 - \frac{1}{2}\e_k)$, for  $\e_k > 0$ arbitrarily small. Let $\s_k: [0,\ell_k] \to M$ be a maximal past directed unit speed timelike geodesic from $\s_k(0) = q_k \in A_k$ to $\s_k(\ell_k) = p_k \in \S_\e$. Since $\s_k$ maximizes the distance to $A_k$, and $V_k \subset J^-(A_k)$, then $\s_k$ also maximizes the distance to $V_k$. Consequently, $V_k$ has no focal points along $\s_k$, except possibly the endpoint $\s_k(\ell_k)$. We may, in fact, push this (potential) focal point into the past by `bending' $V_k$ slightly to the past, keeping $p_k$ fixed.  To carry this out, one can, for example, let $\widehat{V}_k \subset J^-(V_k)$ be a small spacelike paraboloid (in appropriate coordinates near $V_k$) which opens to the past from $q_k \in V_k \cap \widehat{V}_k$. This gives a strict inequality on the corresponding second fundamental forms, and one may apply Proposition 2.3 in \cite{Esch87}, for example, to see that this inequality ensures that the first focal point along $\s_k$, if any, comes strictly later, (further in the past), for $\widehat{V}_k$ than for $V_k$. Furthermore, by taking this paraboloid to be sufficiently flat, (relative to $V_k$), we can ensure that $H_{\widehat{V}_k}(q_k) \ge n (1-\e_k)$.

It follows then that the past normal exponential map $E$ of $\widehat{V}_k$ is a diffeomorphism on some neighborhood of $[0, \ell_k] \times \{p_k\}$, and hence, for some neighborhood $\widetilde{V}_k$ of $p_k$ in $\widehat{V}_k$, the past slice $E(\{t\} \times \widetilde{V}_k)$ is a smooth spacelike hypersurface for all $t \in [0, \ell_k]$. Letting $\theta(t)$ denote the mean curvature of this slice at $\s_k(t)$, the Raychaudhuri equation, together with the curvature condition, give:
$$
\theta'(t) - \dfrac{\theta^2(t)}{n} \ge \mathrm{Ric}(\d_t, \d_t) \ge n  \,.
$$
Using the initial condition, $\theta(0) \ge n (1-\e_k)$, a basic comparison argument gives: $\theta(t) \ge n \tanh(c_k - t)$ for all $t \in [0, \ell_k]$, where $c_k := \tanh^{-1}(1-\e_k)$, (see \cite{Esch87,Grant}). Hence, letting $V'_k := E(\{\ell_k\} \times \widetilde{V}_k)$, then the mean curvature of $V'_k$ satisfies:
$$
H_{V'_k}(p_k) \ge n \tanh(c_k - \ell_k) \ge n \tanh(c_k - \ell_W)   \,.
$$
 
\noindent
Furthermore, for every $x \in V'_k$, we have, $d(x, A_k) \ge d(x, \widehat{V}_k) \ge \ell_k$, by construction. Hence, $V'_k$ cannot meet $I^+(\S_\e)$. Consequently, $V'_k$ serves as a smooth past support hypersurface for $\S_\e$ at $p_k$. But by taking $\e_k$ sufficiently small, we can make $c_k - \ell_W$ arbitrarily large so as to ensure that $H_{V'_k}(p_k) > n(1-\e) = H_{\S_\e}(p_k)$, contradicting the basic second fundamental form inequality $B_{\S_\e}(p_k) \ge B_{V'_k}(p_k)$.\qed

\subsubsection{A Limit  Maximum Principle}

We will use the following notation below. By a \emph{(timelike) diamond neighborhood}, $I_p$, around $p \in M$, we mean a diamond $I_p := I^+(p_-) \cap I^-(p_+)$, for some $p_- << p << p_+$. We denote the corresponding causal diamond by $J_p$, i.e., $J_p := J^+(p_-) \cap J^-(p_+)$. Hence, always $p \in I_p \subset J_p$, so that $p \in \mathrm{int} \, J_p$.

In a globally hyperbolic spacetime, each point admits arbitrarily small causally convex neighborhoods \cite{Penrose}.  Together with the existence of convex normal neighborhoods, this may be used to establish the following.

\begin{lem}\label{dependtrapping} Let $M$ be globally hyperbolic and fix $p \in M$. Then for any neighborhood $U$ of $p$ in $M$, there is a diamond neighborhood $I_p $ of $p$, with $p \in I_p  \subset U$ such that for any achronal set $A \subset\subset I_p$, we have $D(A_p) \subset\subset I_p$. 
\end{lem}

Our present aim is to establish the following `maximum principle' for limit mean curvature:

\begin{lem} [Limit Maximum Principle] \label{limitmax}  Let $(M^{n+1}, g)$ be a globally hyperbolic spacetime satisfying $\mathrm{Ric}(X,X) \ge - n$ for all timelike unit vectors $X$. Let $A_\infty$ and $B_\infty$ be two achronal limits meeting at $p \in A_\infty \cap B_\infty$ such that, near $p$, both achronal limits are acausal, with $B_\infty$ locally to the future of $A_\infty$ (see proof). 
If $A_\infty$ has limit mean curvature $\ge n$ and $B_\infty$ has limit mean curvature $\le n$, then for some neighborhood $U$ of $p$ in $M$, $A_\infty \cap U = B_\infty \cap U$ is a smooth, acausal spacelike hypersurface with $H = n$. 
\end{lem}

\proof Explicitly, we assume that there is a neighborhood $U_0$ of $p$ in $M$ such that $A_\infty \cap U_0$ and $B_\infty \cap U_0$ are acausal, and $B_\infty \cap U_0 \subset J^+(A_\infty \cap U_0)$.   

As with Lemma \ref{limconvex}, the proof involves some careful setup. We first establish a domain of dependence within $U_0$. Let $I_0$ be a diamond neighborhood of $p$ with $p \in I_0 \subset U_0$ satisfying 
Lemma \ref{dependtrapping}. Let $V_0$ be a domain in $A_\8$ around $p$ with $V_0 \subset\subset A_\8 \cap I_0$. Then $D(V_0) \subset\subset  U_0$. 

\vspace{1pc}
\begin{center}
\includegraphics[width=8cm]{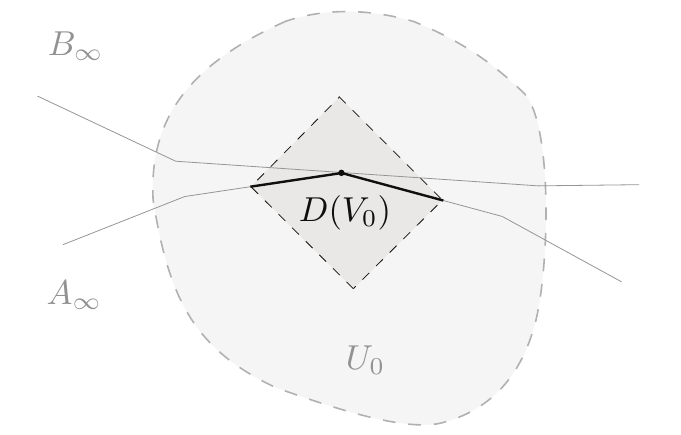}
\end{center}

\vspace{1pc}Since $V_0 \subset A_\8 \cap I_0 \subset A_\8 \cap U_0$, we have that $V_0$ is acausal, and hence $D(V_0)$ is an open, globally hyperbolic subspacetime, with Cauchy surface $V_0$. Let $T$ be a smooth future pointing timelike vector field on $M$.  Let $T'$ be the restriction of $T$ to $D(V_0)$ and set $T_0 := T' / ||T'||_h$, where $h$ is a complete Riemannian metric on $D(V_0)$.  It follows that $T_0$ is a smooth, complete timelike vector field on $D(V_0)$. Hence,  we have a diffeomorphism $\Phi : \field{R} \times V_0 \to D(V_0)$, where, for each $q_0 \in V_0$, the $t$-curve, $\phi_{q_0}(t) = \Phi(t, q_0)$ is the $T_0$-integral curve through $q_0$. This map will be used throughout the proof to relate various (achronal) sets in $D(V_0)$. 

Now let $I_1$ be a diamond around $p$ with 
$p \in I_1 \subset D(V_0)$ satisfying Lemma \ref{dependtrapping}. Let $V_B$ be a small domain around $p$ in $B_\8$, with $V_B$ homeomorphic to an open ball in $\field{R}^n$, and $V_B \subset\subset B_\8 \cap I_1$. Hence, $D(V_B) \subset \subset I_1$. The projection $\pi_2 \circ \Phi^{-1}|_{V_B} : V_B \to V_0$ is continuous, and one-to-one, by achronality. It follows by invariance of domain that its image, $V_A : = \pi_2 \circ \Phi^{-1}|_{V_B}(V_B)$, is a domain in $V_0$ around $p$. By shrinking $V_B$ if necessary, (as a ball), we may suppose also $V_A \subset\subset A_\8 \cap  I_1$. Hence also $D(V_A) \subset \subset I_1$.  

\vspace{1pc}
\begin{center}
\includegraphics[width=9cm]{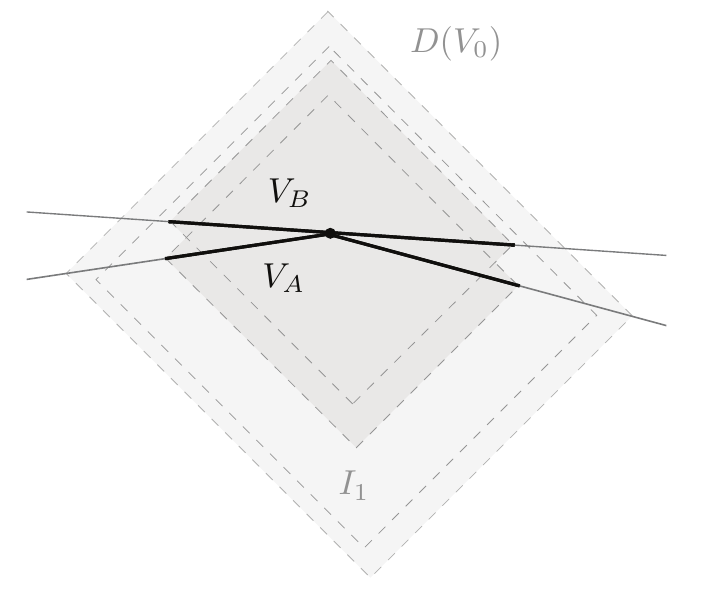}
\end{center}

We emphasize that the points of $V_A$ and $V_B$ are in one-to-one correspondence via the (timelike) integral curves of $T_0$. Hence, fixing any $q \in V_A$, there is a unique point $q' \in V_B$ on the $T_0$-integral curve through $q$ (including the possibility $q' = q$). We will denote this kind of correspondence via the integral curves of $T_0$ by $V_B \approx^{T_0} V_A$, and will use it below on other sets. 

\smallskip
We will show $V_A = V_B$. We have $p \in V_A \cap V_B$. Fix $x \in V_A - \{p\}$. Then, since $V_A$ is homeomorphic to $V_B$, which is homeomorphic to a (hyper)-ball, we can choose a domain $W_A$ in $V_A$ with $\overline{W}_A \subset V_A$ and $x \in \overline{W}_A - W_A =\mathrm{edge} \, W_A$. Let $W_B$ be the corresponding domain in $V_B$, i.e., $W_B \approx^{T_0} W_A$. In fact, since $\overline{W}_A \subset V_A$ and $\overline{W}_B \subset V_B$, we have also $\mathrm{edge} \, W_B \approx^{T_0} \mathrm{edge}\, W_A$. 

\smallskip
Since $D(W_A) \subset D(V_A) \subset \subset J_1$, with $J_1$ globally hyperbolic, it follows that $(W_A, J_1)$ is a `standard data set' as defined by Bartnik in \cite{Bartnik.acta}. Then, since $\overline{W}_A$ is acausal, \cite[Theorem 4.1]{Bartnik.acta} produces a smooth, achronal spacelike hypersurface $\S_A \subset D(W_A)$ of constant mean curvature $H_{\S_A} = n$, with ${\rm edge}\, \S_A = {\rm edge}\, W_A$ and $\S_A \approx^{T_0} W_A$. By Lemma \ref{limconvex}, we have $\S_A \subset J^+(W_A)$. Similarly, now using $(W_B, J_1)$ as the `standard data set', \cite[Theorem 4.1]{Bartnik.acta} and Lemma \ref{limconvex} give a smooth, achronal spacelike hypersurface $\S_B \subset D^-(W_B)$ of constant mean curvature $H_{\S_B} = n$, with ${\rm edge}\, \S_B = {\rm edge}\, W_B$ and $\S_B \approx^{T_0} W_B$. Note that, since ${\rm edge} \, W_A = {\rm edge}\, \S_A$ and ${\rm edge} \, W_B = {\rm edge} \,\S_B$, we have $\mathrm{edge} \, \S_B \approx^{T_0} \mathrm{edge} \, \S_A$. 
 
We now show that $\S_B$ cannot enter $I^-(\S_A)$. Suppose otherwise and let $p_B \in \overline{\S}_B$ and $q_A \in \overline{\S}_A$ such that $\ell = d(\overline{\S}_B, \overline{\S}_A) = d(p_B, q_A) > 0$. By the achronality of $\S_A$ and $\S_B$, and the causal relations on the boundaries, it follows that $p_B \not \in {\rm edge}\, \S_B$ and $q_A \not \in {\rm edge}\, \S_A$, so $p_B \in \S_B$, $q_A \in \S_A$ and $\ell = d(\overline{\S}_B, \overline{\S}_A) = d(\S_B, \S_A) = d(p_B, q_A)$.

\vspace{1pc}
\begin{center}
\includegraphics[width=8cm]{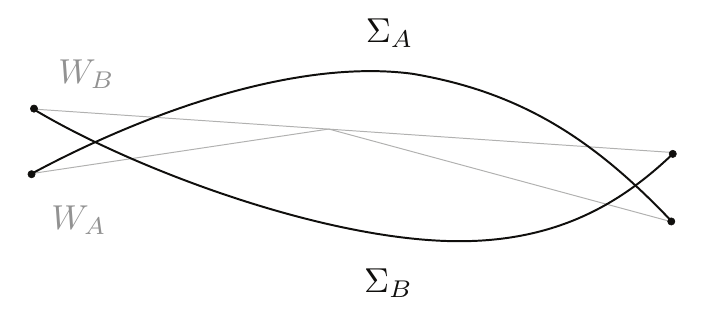}
\end{center}

Hence, the past sphere $S^-_\ell := S^-_\ell(\overline{\S}_A)$ meets $\S_B$ at $p_B \in \S_B \cap S^-_\ell$. Fix an arbitrary intersection point $z \in \S_B \cap S^-_\ell$. Since $S^-_\ell$ is acausal and edgeless, $D(S^-_\ell)$ is an open neighborhood of $z$. If $\S_B$ entered $I^-(S^-_\ell)$, we could produce a timelike curve from $\S_B$ to $S^-_\ell$ of positive length, and hence a curve from $\S_B$ to $\S_A$ of length strictly greater than $\ell$. Hence, $\S_B$ cannot enter $I^-(S^-_\ell)$, and near $z \in \S_B \cap S^-_\ell \subset D(S^-_\ell)$, we have $\S_B$ locally to the future of $S^-_\ell$. Furthermore, for such a $z$, there is some $y \in \overline{\S}_A$ such that $\ell = d(z, \overline{\S}_A) = d(z,y)$. But since $y \in \mathrm{edge\, } \S_A \subset J^-(\mathrm{edge} \, \S_B)$ would lead to a violation of the achronality of $\S_B$, we have $y \in \S_A$. Then, by an argument similar to that in Lemma \ref{limconvex}, (starting with $\S_A$ as a past support hypersurface for itself, bending to the past, and sliding down to $S^-_\ell$), we can show that $S^-_\ell$ has support mean curvature $\ge n$ at $z \in \S_B \cap S^-_\ell$. 
In fact this is true for all points $z' \in S^-_\ell$ near $z$.  
It follows from Theorem~\ref{AGH}, that the intersection $\S_B \cap S^-_{\ell}$ is open in $\S_B$. Since $S^-_{\ell}$ is closed, this intersection is also closed in $\S_B$. Since $\S_B$ is homeomorphic to the (connected) domain $W_B$, $\S_B$ is connected, and hence, $\S_B \cap S^-_{\ell} = \S_B$, i.e., $\S_B \subset S^-_{\ell}$. But since $S^-_{\ell}$ is closed, this implies  ${\rm edge}\, \S_B \subset S^-_{\ell} \subset I^-(\overline{\S}_A)$, which again leads to an achronality violation. 

Hence, $\S_B$ does not meet $I^-(\S_A)$. It follows that $\S_A$ and $\S_B$ are `sandwiched' between $W_A$ and $W_B$, with $p \in \S_A \cap \S_B$, and $\S_ B$ to the future of $\S_A$. 

\vspace{1pc}
\begin{center}
\includegraphics[width=8cm]{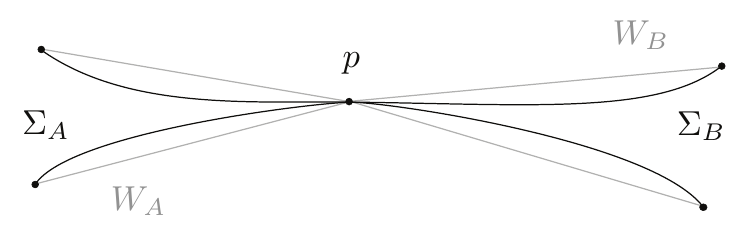}
\end{center}

The (smooth) maximum principle then gives that $\S_A \cap \S_B$ is open in both $\S_A$ and $\S_B$. Suppose $\S_A \cap \S_B$ is not closed in $\S_A$. Then $\S_A \cap \S_B$ has a limit point $p_0 \in \S_A \setminus \S_B$. Then $p_0 \in \overline{\S}_B \setminus \S_B = \mathrm{edge} \, \S_B$. But since $\mathrm{edge} \, \S_B \subset J^+(\mathrm{edge} \, \S_A)$, $p_0 \in \S_A \cap \mathrm{edge} \, \S_B$ leads to an achronality violation. Hence, $\S_A \cap \S_B$ is closed in $\S_A$, and by connectedness, $\S_A \subset \S_B$. By symmetry, we have $\S_A = \S_B$. Hence, ${\rm edge}\, W_A = {\rm edge}\, \S_A = {\rm edge}\, \S_B = {\rm edge}\, W_B$. Thus, the above procedure `sews' $A_\infty$ and $B_\infty$ together along the edges of $W_A$ and $W_B$. 
In particular, we have $x \in {\rm edge}\, W_A = {\rm edge}\, W_B \subset V_B$.  Since $x \in V_A - \{p\}$ was arbitrary, (and since $p \in V_A \cap V_B$), we have $V_A \subset V_B$, and since $V_A \approx^{T_0} V_B$, this means $V_A = V_B$. It follows that $W_A = W_B = \S_B = \S_A$. Hence, near $p$, $A_\infty$ and $B_\infty$ agree and are smooth and spacelike, with mean curvature $n$. \qed

\smallskip
\noindent
{\it Remark:} We note that 
in Lemma \ref{limitmax}, both achronal limits are independently allowed to be future or past limits, with the $A_k$'s and $B_k$'s possibly approaching their limits from the same side, and indeed this is precisely the situation which arises in the proof of Theorem \ref{dSrigid}.

\subsubsection{Proof of Theorem \ref{dSrigid}}

The following lemma establishes the key consequence of the mean curvature assumption \eqref{littleo} imposed on the Cauchy spheres $S^+_k(S)$. 

\begin{lem} \label{asymplem} Let $M^{n+1}$ be a future timelike geodesically complete spacetime satisfying \eqref{EC}. Suppose $S$ is a compact Cauchy surface for $M$ such that each future Cauchy sphere $S^+_k(S)$ has support mean curvature $\ge a_k$, where, letting $n_k := \min\{n, a_k \}$, \eqref{littleo} holds.
Then the Cauchy horosphere $S^-_\8(S)$ has limit mean curvature $\ge n$.
\end{lem}

\smallskip
\noindent
{\it Remark.} For a constant $t$-slice $S_t$ in de Sitter space, one observes that $S^-_{\infty}(S_t) = S_t$.  Hence, the conclusion of this lemma does not hold under the slightly slower asymptotic fall-off 
of \eqref{bigO}.

\proof Let $S_k := S^+_k(S)$ and recall that the sequence of Cauchy prehorospheres is defined by $\widetilde{S}_k := S^-_k(S^+_k(S))$. Fix any $\widetilde{x}_k \in \widetilde{S}_k$. Hence, $d(\widetilde{x}_k, S_k) = k$, and $\widetilde{x}_k$ is joined to some $x_k \in S_k$ by a past-directed $S_k$-maximal unit speed timelike geodesic segment $\alpha_k : [0, k] \to M$, with $\alpha_k(0) = x_k$ and $\alpha_k(k) = \widetilde{x}_k$.

Let $\Sigma_k$ be a smooth past support hypersurface for $S_k$ at $x_k$ with mean curvature $H_{\S_k}(x) \ge a_k - \frac{1}{2}e^{-3k}$. Perturbing $\Sigma_k$ slightly to the past, keeping $x_k$ fixed, as in Lemma \ref{limconvex}, we obtain a smooth past support hypersurface $\widehat{\S}_k$ for $S_k$  at $x_k$  with mean curvature $H_{\widehat{\S}_k}(x) \ge a_k - e^{-3k} \ge n_k - e^{-3k}$, such that $\widehat{\S}_k$ has no focal points along $\a_k$, and hence, such that the past normal exponential map $E$ from $\widehat{\S}_k$ is smooth on $[0, k] \times \widetilde{\S}_k$, for some neighborhood $\widetilde{\S}_k$ of $x_k$ in $\widehat{\S}_k$. Letting $\theta_k(t)$ be the mean curvature of the slice $E(\{t\} \times \widetilde{\S}_k)$ at $\alpha_k(t)$, then $\theta = \theta_k(t)$ satisfies the Raychauduri inequality $\theta^\prime \ge \mathrm{Ric}(\alpha^\prime_k, \alpha^\prime_k) + \theta^2 / n$. By the curvature condition 
this gives $\theta^\prime \ge \theta^2 / n - n$, or, letting 
$\Theta := \theta / n$,
$$ 
\Theta^\prime(t) \ge \Theta^2(t) - 1, \; \; \Theta(0) \ge \dfrac{n_k - e^{-3k}}{n}  \,.
$$ 
Since, for large $k$, $|(n_k - e^{-3k})/n| < 1$, the elementary comparison solution is 
$\tanh (b_k - t)$, with
$$
b_k = \tanh ^{-1}\bigg(\dfrac{n_k -e^{-3k}}{n} \bigg) = \frac{1}{2}\ln\bigg( \dfrac{n+ n_k - e^{-3k}}{n-n_k + e^{-3k}}\bigg) \,.
$$
Thus we have,
\begin{align*}
 \theta_k(k) \; & \ge  \;  n \tanh(b_k - k)\\
 \nonumber \\
& =     \; n \; \dfrac{e^{2b_k} - e^{2k}}{e^{2b_k}+e^{2k}}   \\
 \nonumber \\
 & =    \; n \; \dfrac{(n + n_k - e^{-3k}) - (n - n_k + e^{-3k})e^{2k}}{(n + n_k - e^{-3k}) + (n - n_k + e^{-3k})e^{2k}} \\
 \nonumber \\
& =   \;  n \; \dfrac{(n + n_k - e^{-3k}) - (n - n_k)e^{2k} - e^{-k}}{(n + n_k - e^{-3k}) + (n - n_k)e^{2k}  + e^{-k}}   \; =:   \; \widetilde{\theta}_k  
\end{align*}

\medskip 
Note that using the asymptotic assumption \eqref{littleo}, we have 
$\lim_{k \to \infty} \widetilde{\theta}_k = n$.
Because $\widehat{\S}_k \subset J^-(S_k)$, it follows that the slice $E(\{k\} \times \widetilde{\S}_k)$ is a smooth past support hypersurface for $\widetilde{S}_k$ at $\alpha_k(k) = \widetilde{x}_k$. Since $\widetilde{x}_k$ was arbitrary, we have shown that 
$\widetilde{S}_k$ has mean curvature $\ge \widetilde{\theta}_k$ in the support sense. Since $\widetilde{\theta}_k \to n$, the conclusion follows.\qed

\medskip

The following result is integral to the proof of Theorem \ref{dSrigid}.  It is in some sense analogous to Proposition \ref{hyprigid} and is closely related to \cite[Proposition 3.4]{AndGal}.

\begin{prop} \label{pastsing} Let $M^{n+1}$ be a globally hyperbolic spacetime such that \eqref{EC} holds. Let $S_\8 \subset M$ be a past causally complete achronal limit and suppose that $S_\8$ is acausal with limit mean curvature $\ge n$. Suppose also that $S_\8$ admits a past $S_\8$-ray, $\g$, and let $S^0_\8$ be the connected component of $S_\8$ containing $\g(0)$. Then either $S_\8$ admits a past incomplete timelike $S_\8$-ray, or $S^0_\8$ is a smooth, geodesically complete spacelike past Cauchy surface with mean curvature $H =  n$ and $J^-(S^0_\8)$ splits as:
$$
(J^-(S^0_\8),g) \approx ( (-\8,0] \times S^0_\8, -dt^2 + e^{2t}h),
$$
where $h$ denotes the induced metric on $S^0_\8$. 
\end{prop}

\proof Observe that, since $S_\8$ is acausal, every $S_\8$-ray, future or past, is timelike. Assume all past $S_\8$-rays are past complete. Hence $\g$ is timelike and past complete. Consider  the future horosphere $S^+_\8(\g)$. By Lemma \ref{limithoro}, $S^+_\8(\g)$ has limit mean curvature $\le  n$. Let $S^+$ be the connected component of $S^+_\8(\g)$ containing 
$\g(0)$. Hence, the intersection $S^0_\8 \cap S^+$ is nonempty and closed. Since $S^0_\8$ is acausal, it follows (using the argument of Theorem \ref{rays} locally)
that $S^+$ must also be locally acausal and to the future of $S^0_\8$ near any intersection point $x \in S^0_\8 \cap S^+$.  
Hence, using Lemma \ref{limitmax}, $S^0_\8 = S^+$ is a smooth spacelike hypersurface with mean curvature $H =  n$.   Thus, by a trivial modification of the proof of
\cite[Proposition 3.4]{AndGal}, we have that the normal past  $N^-(S^0_\8)$ (generated  by the past directed $S^0_\8$-normal geodesics) splits as $((-\8, 0] \times S^0_\8, -dt^2 + e^{-2 t}h)$. Then by adapting the proof of Theorem 3.68 in \cite{BEE}, using the past causal completeness of $S^0_\8$ and the warped product structure, we get that $S^0_\8$ is geodesically complete. Hence, it follows from Theorem 3.69 in \cite{BEE} that $N^-(S^0_\8) = J^-(S^0_\8)$.\qed

\bigskip  
We now proceed to the proof of Theorem \ref{dSrigid}.
\proof[Proof of Theorem \ref{dSrigid}]   Since $S^-_\8(S)$ is inherently future bounded by $S$, it is acausal and all $S^-_\8(S)$-rays, future or past, are timelike. We will suppose every past $S^-_\8$-ray is complete and show (2) in the statement of Theorem \ref{dSrigid} holds. By Lemma \ref{asymplem}, $S^-_\8(S)$ has limit mean curvature $\ge n$. Then, letting $S^-$ be the connected component of $S^-_\8(S)$ which contains $\g(0)$, Proposition~\ref{pastsing} gives that $S^-$ is a smooth, geodesically complete, spacelike past Cauchy surface, with mean curvature $H = n$, and gives the isometry
$$
(J^-(S^-),g) \approx ((-\8,0] \times S^-, -dt^2 + e^{2t}h)  \,.
$$

Note that the future radial rays from $S^-_\8$ are all timelike and future complete. Since $S^-$ is smooth, there must only be one such ray from each point $p \in S^-$, and it must be the future normal geodesic from $p \in S^-$. Hence, the future normal exponential map $E$ is a diffeomorphism onto the future image $N^+(S^-) = E([0, \8) \times S^-)$. The standard comparison argument via the Raychaudhuri equation gives $H_t \le n$  for the future normal slice $N_t := E(\{t\} \times S^-)$ (see e.g. \cite[Theorem 7]{Grant}). But the usual argument does not give $H_t = n$. To get the splitting to the future, we will instead identify $N_t$ with a portion of (what is essentially) the Cauchy horosphere associated to the Cauchy surface $S_t :=S^+_t(S)$. Like $S^-_\8$, this horosphere will inherit limit mean curvature $\ge n$ from the sequence $\{S^+_k(S)\}$, and we can run our arguments again to get $H = n$ for this horosphere, (locally), and hence $H_t = n$ for the slice $N_t$.

Fix $t > 0$. As in (the time dual of) Lemma \ref{equidist}, we have $S^+_k(S) = S^+_{k-t}(S^+_t(S)) = S^+_{k-t}(S_t)$, and hence, $S^-_{k-t}(S^+_k(S)) = S^-_{k-t}(S^+_{k-t}(S_t))$. The same monotonicity argument for the usual Cauchy prehorospheres shows that the sequence $\{J^-(S^-_{k-t}(S^+_k(S)))\} = \{J^-(S^-_{k-t}(S^+_{k-t}(S_t)))\}$ is decreasing. Letting $\widetilde{S}_{k-t} := S^-_{k-t}(S^+_k(S))$, consider the horosphere
$$
S^-_{\8 - t} : = \d \bigg( \bigcap_k J^-(\widetilde{S}_{k-t}) \bigg)   \,.
$$

\smallskip
\noindent
We want to show $N_t \subset S^-_{\8-t}$. We first note that, as with the usual prehorospheres, $\widetilde{S}_{k-t} = S^-_{k-t}(S^+_{k-t}(S_t))$ is future bounded by $S_t$. Let $x_\8 \in S^- \subset S^-_\8$ and fix a sequence $x_k \in \widetilde{S}_k$ with $x_k \to x_\8$. Since $x_k \in \widetilde{S}_k = S^-_k(S^+_k(S))$, there is a future maximal unit speed timelike geodesic segment, $\s_k : [0, k] \to M$, joining $\s_k(0) = x_k$ to $\s_k(k) \in S^+_k(S)$. Then $x_{k-t} := \s_k(t) \in S^-_{k-t}(S^+_k(S)) = \widetilde{S}_{k-t}$. Letting $x_{-1} \in I^-(x_\8)$, we have $x_{k-t} \in J^+(x_{-1}) \cap J^-(S_t)$, for large $k$. Hence, by passing to a subsequence if necessary, the sequence $\{x_{k-t}\}$ has a limit, $x_{\8-t}$, which must be contained in $S^-_{\8 - t}$, by Proposition \ref{sequential}. Since $S^-_{\8 -t}$ is future bounded, it admits a timelike future $S^-_{\8 -t}$-ray $\eta$ from $x_{\8 -t}$. Since $t = d(\s_k(0), \s_k(t)) = d(x_k, x_{k-t}) \to d(x_\8, x_{\8-t})$,
there is a maximal unit speed geodesic segment $\b : [0, t] \to M$ from $x_\8$ to $x_{\8-t}$. Finally, since $d(\widetilde{S}_k, \widetilde{S}_{k-t}) = t$, we have $d(S^-_\8, S^-_{\8-t}) = t$. It follows that the concatenation $\s = \b + \eta$ is an $S^-_\8$-ray from $x_\8 \in S^-$. Since $\s$ is also an $S^-$-ray, paramterizing $\s$ as a unit speed geodesic, we have $\s(t) = \b(t) = x_{\8-t}$.  This shows $N_t \subset S^-_{\8-t}$.

Replacing $e^{2k}$ by $e^{2(k-t)} = e^{2k-2t}$ in the calculation in Lemma \ref{asymplem}, that is, sliding the past support hypersurface for $S^+_k(S)$ down for a time $k -t$ instead of $k$, shows that $S^-_{\8 -t}$ has limit mean curvature $\ge n$. Recall that $N_t \subset S^-_{\8-t}$ has (smooth) mean curvature $H_t \le n$. Hence, working locally, and viewing $N_t$ as the (constant) achronal limit of itself, Lemma \ref{limitmax} gives that $N_t$ has constant mean curvature $H_t = n$. Since $t > 0$ was arbitrary, all future normal slices have constant mean curvature $H = n$. Plugging this back into the Raychaudhuri equation, the characterization of the equality case gives that each slice $N_t$ is totally umbilic with $B_t = h_t$, where $h_t$ is the induced metric on $N_t$.  From this it easily follows that $N^+(S^-) \approx ([0, \8) \times S^-, -dt^2 + e^{2t}h)$.   As in Remark 3.71 of \cite{BEE}, and the related discussion, which cites also \cite{Powell}, this warped product structure means that $N^+(S^-)$ is future null and timelike geodesically complete. Hence, any future causal geodesic starting from $S^-$ can never leave $N^+(S^-)$. Since any $y \in J^+(S^-)$ is joined to some $s \in S^-$ by a future causal geodesic segment from $s \in S^-$, we have $y \in N^+(S^-)$. Hence, $J^+(S^-) = N^+(S^-)$, and $J(S^-) = N(S^-)$, with
$$
(J(S^-),g) \approx ((-\8,\8) \times S^-, -dt^2 + e^{2t}h)  \,.
$$
In particular, $H^+(S^-) \subset J^+(S^-) = N^+(S^-)$, but by Theorem 3.69 in \cite{BEE}, $S^-$ is a Cauchy surface for $N(S^-) = J(S^-)$. Hence, $H^+(S^-) = \emptyset$. Recalling that also $H^-(S^-) = \emptyset$, we have that $S^-$ is a Cauchy surface for $M$. By achronality, this means $S^-_\8 = S^-$, which gives the conclusion. \qed

\bigskip

\noindent
{\it Acknowledgement.} We thank the referee for a careful reading of the paper and for numerous helpful suggestions.

%

\section{Appendix}

The following is part of  Definition 3.3 in \cite{AGH}:

\begin{Def} [Support Mean Curvature with One-Sided Hessian Bounds] \label{defHessbnds} Let $S$ be a $C^0$ spacelike hypersurface in a spacetime $M$ and $a \in \bbR$. 
We say $S$ has \emph{support mean curvature $\ge a$ with one-sided Hessian bounds} if, fixing any compact subset $K \subset S$, there is a compact set $\widehat{K} \subset TM$ and a constant $C_K > 0$ such that for all $q \in K$ and all $\e > 0$, there is a $C^2$ past support hypersurface $S_{q, \e}$ for $S$ at $q$ such that

\ben
\vspace{-.5pc}
\item [i)] The future unit normal field, $\eta_{q,\e}$, of $S_{q,\e}$ satisfies: $\eta_{q,\e}(q) \in \widehat{K}$
 
\vspace{-.5pc}
\item [ii)] The mean curvature, $H_{q, \e}$, of $S_{q, \e}$ satisfies: $H_{q,\e}(q) \ge a - \e$

\vspace{-.5pc}
\item [iii)] The second fundamental form, $B_{q, \e}$, of $S_{q, \e}$ satisfies: $B_{q,\e}(q) \ge - C_K $
\een
\end{Def}
The ``one-sided Hessian bounds" refers to condition (iii), which requires the  second fundamental forms associated to the family of support hypersurfaces to be locally uniformly bounded below, as specified.  As discussed in \cite{AGH}, this condition insures the uniform ellipticity of the mean curvature operator with respect to the family of support  hypersurfaces.  

Proposition 3.5 in \cite{AGH} guarantees that when the support surfaces $S_{q, \e}$ are smooth past point spheres, the one-sided bound on the second fundamental forms holds provided that the set of  support normals is {\it locally compact} in the following sense:  A set of vectors $\mathcal{X} \subset TM$ is locally compact if, over any compact $K \subset M$, the subset $\mathcal{X} \cap \pi^{-1}(K)$ is compact, where $\pi : T(M) \to M$ is the natural projection.

\begin{prop}\label{horosuppnormals} Let $M$ be a globally hyperbolic spacetime and suppose that $S^-_\8$ is a past horosphere such that all future $S^-_\8$-rays are timelike and future complete. Let $\mathcal{N}$ be the set of the initial tangent vectors of all future $S^-_\8$-rays, parameterized as unit speed geodesics. Then $\mathcal{N}$ is locally compact.
\end{prop}

A detailed proof of this is given in \cite[Section 5.1]{Vega}.  The argument is essentially as follows.  If $\mathcal{N}$ is not locally compact then there is a sequence of initial directions which approaches a null direction.  Since the limit of 
$S^-_\8$-rays is an $S^-_\8$-ray, the null geodesic in the direction of this limit null direction will be an $S^-_\8$-ray, contradicting the fact that all $S^-_\8$-rays are timelike.


\providecommand{\bysame}{\leavevmode\hbox to3em{\hrulefill}\thinspace}
\providecommand{\MR}{\relax\ifhmode\unskip\space\fi MR }
\providecommand{\MRhref}[2]{%
  \href{http://www.ams.org/mathscinet-getitem?mr=#1}{#2}
}
\providecommand{\href}[2]{#2}

\end{document}